\documentclass[twocolumn]{aastex701}

\usepackage{amsmath}
\usepackage{CJK}
\usepackage{comment}
\usepackage{graphicx}
\usepackage{natbib}
\usepackage{multirow}
\usepackage{upgreek}

\begin{document}

\title{Characterizing the Time Variability of 2M1207~A+b with JWST NIRSpec/PRISM}

\author[0000-0002-7139-3695]{Arthur D. Adams}
\email[show]{arthur@virginia.edu}
\affiliation{Department of Astronomy, University of Virginia, 530 McCormick Road, Charlottesville, VA 22904, USA}

\author[0000-0003-2969-6040]{Yifan Zhou}
\email{yzhou@virginia.edu}
\affiliation{Department of Astronomy, University of Virginia, 530 McCormick Road, Charlottesville, VA 22904, USA}

\author[0000-0002-2919-7500]{Gabriel Dominique-Marleau}
\email{gabriel.marleau@uni-due.de}
\affiliation{Division of Space Research and Planetary Sciences, Physics Institute, University of Bern, Sidlerstr.~5, 3012 Bern, Switzerland}
\affiliation{Max-Planck-Institut f\"ur Astronomie, K\"onigstuhl 17, 69117 Heidelberg, Germany}
\affiliation{Fakult\"at f\"ur Physik, Universit\"at Duisburg--Essen, Lotharstraße 1, 47057 Duisburg, Germany}

\author[0000-0003-3714-5855]{Daniel Apai}
\email{apai@arizona.edu}
\affiliation{University of Arizona, 933 N.~Cherry Avenue, Tucson, AZ 85721, USA}

\author[0000-0003-4614-7035]{Beth A. Biller}
\email{bb@roe.ac.uk}
\affiliation{Scottish Universities Physics Alliance, Institute for Astronomy, University of Edinburgh, Blackford Hill, Edinburgh EH9 3HJ, UK}
\affiliation{Centre for Exoplanet Science, University of Edinburgh, Edinburgh EH9 3HJ, UK}

\author[0000-0001-5365-4815]{Aarynn L. Carter}
\email{aacarter@stsci.edu}
\affiliation{Space Telescope Science Institute, Baltimore, MD 21218, USA}

\author[0000-0003-0489-1528]{Johanna M. Vos}
\email{johanna.vos@tcd.ie}
\affiliation{School of Physics, Trinity College Dublin, The University of Dublin, Dublin 2, Ireland}
\affiliation{Department of Astrophysics, American Museum of Natural History, Central Park West at 79th Street, New York, NY 10034, USA}

\author[0000-0001-8818-1544]{Niall Whiteford}
\email{nwhiteford@amnh.org}
\affiliation{Department of Astrophysics, American Museum of Natural History, Central Park West at 79th Street, New York, NY 10034, USA}

\author[0000-0001-7058-1726]{Stephan Birkmann}
\email{Stephan.Birkmann@esa.int}
\affiliation{European Space Agency, European Space Astronomy Centre, Camino Bajo del Castillo s/n, E-28692 Villanueva de la Ca\~{n}ada, Madrid, Spain}

\author[0000-0001-7356-6652]{Theodora Karalidi}
\email{tkaralidi@ucf.edu}
\affiliation{University of Central Florida, Physical Sciences Bldg. 430, 4111 Libra Drive, Orlando, FL 32816, USA}

\author[0000-0003-2278-6932]{Xianyu Tan}
\email{xianyut@sjtu.edu.cn}
\affiliation{Tsung-Dao Lee Institute \& School of Physics and Astronomy, Shanghai Jiao Tong University, Shanghai 201210, People's Republic of China}

\author[0000-0003-0774-6502]{Jason Wang}
\email{jason.wang@northwestern.edu}
\affiliation{Northwestern University, 1800 Sherman Ave, Evanston, IL 60201, USA}

\author[0000-0003-0568-9225]{Yuhiko Aoyama}
\email{aoyama@sysu.edu.cn}
\affiliation{Sun Yat-sen University, Zhuhai 519082, People's Republic of China}

\author[0000-0003-2649-2288]{Brendan P. Bowler}
\email{bpbowler@ucsb.edu}
\affiliation{Department of Physics, University of California, Santa Barbara, Santa Barbara, CA 93106, USA}

\author[0000-0001-5579-5339]{Mickaël Bonnefoy}
\email{bonnefmi@gmail.com}
\affiliation{Univ. Grenoble Alpes, CNRS, IPAG, F-38000 Grenoble, France}

\author[0000-0002-3053-3575]{Jun Hashimoto}
\email{jun.hashimto@nao.ac.jp}
\affiliation{National Institutes of Natural Sciences, Mitaka, Tokyo 181-8588, Japan}

\begin{abstract}

We present JWST NIRSpec/PRISM IFU time-resolved observations of 2M1207~A and b (TWA 27), a $\sim 10$ Myr binary system consisting of a $\sim 2500$ K sub-stellar primary hosting a $\sim 1300$ K companion. Our data provide 20 time-resolved spectra over an observation spanning 12.56 hours. We provide an empirical characterization for the spectra of both objects across time. For 2M1207~A, non-linear trend models are statistically favored within the ranges 0.6--2.3 $\upmu$m and 3.8--5.3 $\upmu$m. However, most of the periods constrained from sinusoidal models exceed the observing window, setting a lower limit of 12.56 hours. We find the data at H$\alpha$ and beyond 4.35~$\upmu$m show a moderate time correlation, as well as a pair of light curves at 0.73--0.80~$\upmu$m and 3.36--3.38~$\upmu$m. For 2M1207~b, light curves integrated across 0.86--1.77 $\upmu$m and 3.29--4.34 $\upmu$m support linear trend models. Following the interpretation of \citet{Zhang2025}, we model the 2M1207~b data with two 1D atmospheric components, both with silicate and iron condensates. The model of time variability as changes to the cloud filling factor shows broad consistency with the variability amplitudes derived from our data. Our amplitudes, however, disagree with the models at $\approx$0.86--1 $\upmu$m. While an additional model component such as rainout chemistry may be considered here, our analysis is limited by a low signal-to-noise ratio. Our results demonstrate the capability of JWST to simultaneously monitor the spectral variability of a planetary-mass companion and host at low contrast.

\end{abstract}


\section{Introduction}\label{sec:introduction}

Observations of planetary-mass objects with the James Webb Space Telescope (JWST) are revealing the complexity of their atmospheres with unprecedented detail \citep[e.g.][]{Miles2023,Dyrek2024,Powell2024,Bell2024,Hu2024,McCarthy2025}. We are now able to place increasingly precise constraints on the thermal and chemical structures of directly-imaged exoplanets and brown dwarf companions. Direct imaging observations are especially feasible for systems with warm (L and T spectral type) companions on wide orbits (projected separations $\gtrsim 1^{\prime\prime}$). Leveraging the observational viability of these companions with atmospheric modeling can provide evidence for hypotheses of how they form and evolve, for example by comparing chemical abundance ratios predicted for objects which form as a binary with their host versus in their proto-planetary disk \citep[e.g.][]{Molliere2020,Xuan2022,Adams2023,Gaarn2023,Inglis2024,Nasedkin2024b}.

The warmth and brightness of these planetary-mass companions reflects their youth ($\sim$1--100 million years), which comes with additional peculiar properties that provide observable probes of planetary evolution. For one, young companions rotate rapidly \citep[with periods of order hours to tens of hours; see e.g.][]{Zhou2016,Bryan2020b}, a consequence of retaining much of the initial angular momentum from formation. Secondly, the combination of low mass and high temperature produces a low surface gravity, which allows clouds to condense in their photospheres at a wider range of effective temperatures than older, higher-mass objects of similar spectral types \citep[e.g.][]{Madhusudhan2011c,Skemer2011,Skemer2014,Marley2012,Faherty2016}. Thirdly, at ages of 1--10 Myr, systems still harbor a disk, which provides a material source for magnetospheric accretion onto the star \citep[see e.g.][]{Hartmann2016}. Disks have been detected down to sub-stellar \citep[e.g.][]{Luhman2005,Joergens2013,Bayo2017,Scholz2023,Seo2025} and planetary massses \citep[e.g.][]{Bowler2011,Isella2019,Benisty2021}. This material has the potential to accrete onto planetary-mass companions in the early stages of their development.

\subsection{Variability in Substellar Atmospheres}\label{sec:introduction:variability}
The combination of rapid rotation, cloudy atmospheres, and the potential for active accretion mean that the spectra of planetary-mass companions can often be highly variable. Theories for time variability in sub-stellar spectra have been summarized in works such as \citet{Apai2017,Biller2017,Artigau2018,Sánchez-Lavega2023,Lee2024,Fuda2024b}. L and T dwarfs are cool enough to exhibit cloudy photospheres and therefore can exhibit time variability through inhomogeneous cloud distributions that rotate in and out of view. Variability can occur through variations in the thicknesses of cloud layers and their relative covering fraction of the observed hemisphere \citep[see e.g.][]{Apai2013,Morley2014,Zhang2025}. Fluctuations in the vertical temperature structure have been employed to explain the phase shifts of time variability with wavelength \citep{Robinson2014}. This source of these possible vertical fluctuations was linked to theories about thermo-chemical instabilities originally proposed in \citet{Tremblin2015,Tremblin2016}. More recently, works such as \citet{Apai2017,Fuda2024b} provide a more comprehensive explanation for variability over longer baselines, using planet-wide bands with spatially periodic brightness variations (referred to as ``planetary-scale waves''). Young brown dwarfs can be warm enough to exhibit M spectral types, for whom starspots are another source of variability, driven through rotational modulation \citep[e.g.][]{Davenport2015,JRBarnes2015,Wanderley2023,Mori2024}. Finally, accretion of disk material is known to cause time variability \citep[][]{Demars2023} which manifests in the variation in the strength and width of accretion-sensitive emission lines, including H$\alpha$. Therefore, studying variability for young directly-imaged planetary-mass companions is both an important and feasible probe of their atmospheric structures.

These theoretical sources of time variability have been tested through observations with several generations of telescopes, most recently with JWST \citep{Gardner2023}. Spot models were originally employed to describe photometric variability modulations of L/T dwarfs as well as M dwarfs; one example is the brown dwarf binary WISE1049~AB \citep[also known as Luhman 16, see][]{Luhman2013,Karalidi2016}. Follow-up observations of this system with the Hubble Space Telescope (HST) across $\approx$100~rotation periods have demonstrated that the inclusion of planetary-scale waves (also referred to as bands) provides the most consistent match to long-term variability \citep{Fuda2024a}. This study is the latest in a series of variability monitoring programs with HST \citep[see also][]{Apai2017,Lew2016,Lew2020,Zhou2016,Zhou2018,Zhou2020,Manjavacas2019a,Manjavacas2019b,Miles-Paez2019}. Along with the Spitzer Space Telescope (hereafter Spitzer) and the Transiting Exoplanet Survey Satellite (TESS), these have been the primary observatories for the characterization of variable sub-stellar objects prior to the launch of JWST \citep[see also e.g.][]{Zapatero2003,Artigau2009,Gillon2013,Radigan2012,Radigan2014a,Metchev2015,Karalidi2016,Vos2022,McCarthy2024a}. JWST now provides high signal-to-noise, spectrally-resolved variability data in the near-infrared, extending the legacy of insights into the chemical and physical nature of sub-stellar variability. This has made possible, for example, the study of the aforementioned WISE1049~AB system using MIRI/LRS \citep{Wright2023} and NIRSpec/PRISM \citep{Jakobsen2022} observations \citep[GO program~2965, P.I.~Biller; see][]{Biller2024,Chen2024}, as well as NIRSpec and MIRI time-series observations of SIMP 0136+0933 \citep[GO program~3548, P.I.~Vos; see][]{McCarthy2025}. At a mass ratio and effective temperatures similar to the subject of this study, the VHS 1256 system has seen multiple variability studies prior to the launch of JWST \citep{Bowler2020,Zhou2020a,Zhou2022b}. These studies reveal that time variability is a complex phenomenon resulting from a combination of clouds, hot spots, chemical variation, and magnetically-driven features such as auror\ae\ and accretion. These also lay the contextual foundation for new observations of one of the most well-studied systems hosting a planetary-mass companion: \objectname[TWA 27]{2MASS J12073346$-$3932539} (hereafter 2M1207), alternatively known as TWA~27.

\subsection{An Overview of the 2M1207 System}\label{sec:introduction:2M1207}
2M1207~b was the first planetary-mass companion to be directly imaged \citep{Chauvin2004,Chauvin2005a,Song2006}. Prior to this study, the most recently published data on this system were JWST NIRSpec IFU observations taken on 2023 February 7 as part of GTO program 1270 \citep[P.I.~Birkmann, originally published in][]{Luhman2023}. These observations use three filter/disperser pairs (G140H/F100LP, G235H/F170LP, and G395H/F290LP) spanning 0.97--5.27~$\upmu$m at a mean spectral resolution $R {\sim}2700$. \citet{Manjavacas2024} fit the spectra of both objects using grids of cloudy \citep[\texttt{BT-Settl}, see][]{Allard2012} and cloud-free \citep[\texttt{ATMO}, see][]{Tremblin2015,Tremblin2016} atmospheric models. They report effective temperatures of approximately 2600 and 1300~K and gravities of $\log g=4.0\pm0.5$ and 3.5--3.6, respectively, as constrained by model grids. More recently, \citet{Zhang2025} fit the data of 2M1207~b with a series of atmospheric models and place precise constraints on the cloud properties which form a reference point for our analysis (see \S \ref{sec:atmosphere-models}).

Constraining both the rotation and potential presence of accreting material are important in characterizing the variability of systems such as 2M1207. \citet{Zhou2016} find a best-fit period of $10.7^{+1.2}_{-0.6}$ hours for 2M1207~b when applying a sinusoidal model to HST variability monitoring data. Concerning accretion, \citet{Mohanty2007} originally found that modeling of near-infrared VLT (Very Large Telescope) data of 2M1207~b preferred the inclusion of a gray opacity source, providing the first evidence of an observable contribution from circum-planetary disk material. The disk has been imaged with ALMA \citep{Ricci2017}, adding 2M1207 to the growing catalog of planetary systems imaged at spatial scales of just a few~au \citep[e.g.][]{Andrews2018,Huang2018,Cieza2021,Jennings2022,ZhangS2023,Teague2025}. The system also has evidence for an outflow that is oriented close to the plane of the sky, and may be perpendicular to the disk \citep{Whelan2007,Whelan2012}. Asymmetries in disk structures have been observed to contribute to time variability in the near-infrared through observing programs such as the Spitzer Young Stellar Object Variability (YSOVAR) \citep[see e.g.][]{Faesi2012,Poppenhaeger2015,Kesseli2016,Meng2016}. \citet{Luhman2023} show evidence for accretion in the JWST spectrum of 2M1207~b through the detection and characterization of Paschen and atomic helium lines, which trace accretion in young objects \citep[see e.g.][]{Natta2004,Edwards2006}. \citet{Marleau2024} use these data along with HST WFC3/UVIS2 data (program ID: 12225, P.I.~Reiners) to estimate the mass accretion rate of 2M1207~b as $10^{-13}$--$5\times10^{-12}$ M$_\odot$ yr$^{-1}$. The HST data show a signal of H$\alpha$ emission, though at only 2.2$\sigma$ this was reported as a non-detection. They measure an intrinsic width of $67\pm9$ km s$^{-1}$ for three of the detected lines, $\approx 60$\%\ of the free-fall velocity of 2M1207~b, which lends evidence to the interpretation that these lines trace accreting material. Most recently, mid-infrared data of 2M1207~b from JWST MIRI show silicate absorption at 8--10 $\upmu$m, and an infrared excess from imaging at 15 $\upmu$m \citep{Patapis2025}. For the primary, \citet{France2010} use UV spectra from HST to provide evidence that the disk material exhibits signatures of hot gas that are consistent with accretion shocks. \citet{Venuti2019} estimate the mass accretion rate at $\sim 10^{-11}$ M$_\odot$ yr$^{-1}$ for the primary, 2M1207~A, using X-shooter spectra. \citet{Ricci2017} image the circum-stellar disk material around 2M1207~A using ALMA but do not detect disk emission around 2M1207~b. With JWST, \citet{Manjavacas2024} find their 2M1207~A data diverges from atmospheric model fits beyond $\approx 2.5$~$\upmu$m, which the authors attribute to infrared excess from circumstellar disk material.

Building upon this understanding of the rotational and accretion properties of the 2M1207 system, we present new time-resolved JWST data along with an initial characterization of the spectral variability of 2M1207~A and b. The observations are described in \S~\ref{sec:observations}, and then the spectral and temporal features of the data are quantified in \S~\ref{sec:variability-models}. We are able to interpret our data on 2M1207~b in the context of a recent atmospheric retrieval on previous JWST data; this is described in \S~\ref{sec:atmosphere-models}. We discuss possible interpretations of the variability seen in the data of 2M1207~A and b in \S~\ref{sec:discussion}; our results are summarized in \S~\ref{sec:conclusions}.

\section{Observations and Data Reduction}\label{sec:observations}

\subsection{Observations}\label{sec:observations:observing_details}
We observed 2M1207~A+b using the JWST NIRSpec IFU on 2024 June 29--30 using the PRISM/CLEAR configuration (GO Program 3181, P.I.~Zhou R$\sim$30--300, 0.6--5.3~$\upmu$m). The data are available on MAST at \dataset[DOI 10.17909/0yte-e163]{\doi{10.17909/0yte-e163}}. The observations consist of a 12.56-hour time series, designed to cover one full rotation period of 2M1207~b as originally measured in \citet{Zhou2016}. The target system (separation = 0.78\arcsec, $\Delta J = 7.0$~mag) was monitored in 20 consecutive exposures, each using 10 groups in NRSIRS2RAPID readout mode with two integrations and a four-point dither pattern, resulting in 1284~s of integration time per exposure. We placed the mid-point of the binary at the center of the detector, and kept both components within the $3'' \times 3''$ NIRSpec/IFU field of view.

Reference star observations of TWA-28 (M8.5, $J = 13.03$~mag) were obtained at both the beginning and end of the time series to estimate the point-spread function (PSF), enable reference star differential imaging, and to monitor PSF stability. The reference observations used identical instrument configurations but with one integration per exposure. The entire observing sequence was executed without interruption to maintain consistent and stable PSF properties.

\subsection{Initial Data Reduction}\label{sec:observations:initial_data_reduction}
\begin{figure*}
    \centering
    \includegraphics[width=\linewidth]{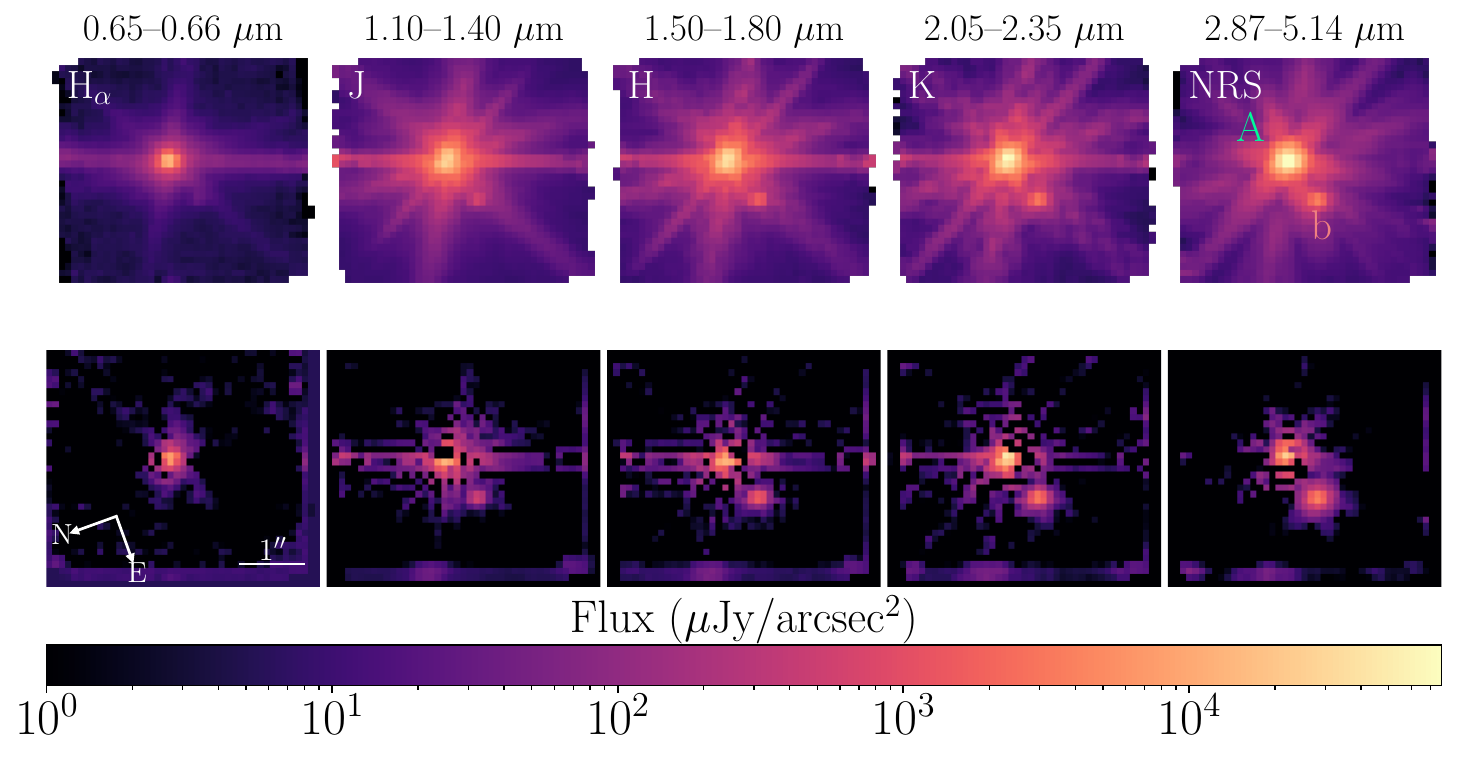}
    \caption{Wavelength-binned images of the 2M1207 system (also known as TWA 27). The top and bottom rows show NIRSpec IFU PRISM/CLEAR (0.6--5.3~$\upmu$m, R$\sim$30--300) images before and after the PSF subtraction, respectively. The field of view of the IFU is $3\times3^{\prime\prime}$. All images are plotted on the same logarithmic intensity-to-color scale. We choose to bin the images according to approximations of common band boundaries: H$_\alpha$, $J$, $H$, $K$, and the total range of NIRSpec NRS when used with the G395M/H dispersers. This highlights, for example, the very red color \citep[$J$--$K$ = 2.72~mag;][]{Luhman2023} of 2M1207~b as a planetary-mass companion compared with field objects of a similar late L spectral type.}
    \label{fig:primary_subtraction}
\end{figure*}

The data reduction process began with \texttt{uncal} files retrieved from the MAST archive. The data were originally processed using JWST Calibration References Data System (CRDS) context \texttt{jwst\_1242.pmap}. The results presented for 2M1207~b use this context. We reprocessed the data using the latest CRDS context \texttt{jwst\_1377.pmap}, \texttt{SDP\_VER=2024\_1a} and verified that the results are generally consistent. One exception is the presence of time-isolated non-astrophysical outliers in 2M1207~A's data which motivated the use of the latest context for 2M1207~A's results. We processed these files in three stages. Stage 1 applied detector-level corrections including enhanced snowball rejection optimized for NIRSpec, reference pixel correction, saturation flagging, and ramp fitting to convert raw readouts into calibrated count rates. Stage 2 transformed these count rates into calibrated spectroscopic products by applying micro-shutter array (MSA) flagging, flat-fielding, path loss corrections, and photometric calibration. Stage 3 constructed preliminary 3D spectral cubes using drizzle weighting in the IFU alignment coordinate system using the drizzle weighting method with a kernel size of 7. This step also included running the \texttt{outlier\_detection} routine to account for spurious outliers such as those from cosmic rays. This procedure produced 20 spectral cubes for the science target and two for the reference star.

The PSF subtraction is performed by subtracting the reference star (TWA~28) from the target. We created simulated PSFs using WebbPSF to constrain the center of both the reference and target star. These PSFs match the observations to reproduce the instrumental effects. Figure~\ref{fig:primary_subtraction} shows a sample of IFU images integrated across typical infrared band ranges, both pre- and post- PSF subtraction. We created pixel masks to exclude regions that could interfere with centroid determination. These masks exclude bright central pixels of 2M1207~A and a $3\times3$ box around 2M1207~b. For centroid determination, we used a multi-stage PSF fitting approach to separate position finding from flux measurement. We first fit each wavelength slice to find approximate positions. We then fit these positions with a polynomial across wavelengths to capture their systematic variation. The final step fixed these polynomial-fitted positions and focused solely on precise flux determination of the primary at each wavelength. For PSF subtraction, we used reference differential imaging. This method scales the reference star to match the primary star's PSF before subtraction. This scale factor excludes a 3-pixel radius region around 2M1207~b in order to mitigate flux loss. The procedure delivered primary-subtracted IFU cubes. We determined the host's position (precision $\sim 0.03$ pixel) and spectrum from the best-fitting synthetic PSF model fluxes. While the reference star and 2M1207~A are similar in spectral type and brightness, they are not identical, and some systematic residuals remain, as seen in the central panels of the lower row of Figure~\ref{fig:primary_subtraction}.

We determined the companion's position by fitting a PSF model to the primary-subtracted frames. We used the same multi-stage approach that we applied to the host star. We then extracted the companion's spectrum using the \texttt{jwst} pipeline's \texttt{Extract1dStep} function. We placed the extraction aperture at the best-fitting position. We also extracted spectra from several positions at the same separation as 2M1207~b. This allowed us to evaluate the level of contamination. This procedure delivered both photometric and astrometric measurements, including a new measurement of the position angle and separation of 2M1207~b. We include a brief analysis, including new orbital constraints, in Appendix \ref{sec:appendix:astrometry}.

\subsection{Estimating Post-Subtraction Residuals}\label{sec:observations:post_subtraction_residuals}
\begin{figure}
\begin{center}
\includegraphics[width=6.5cm]{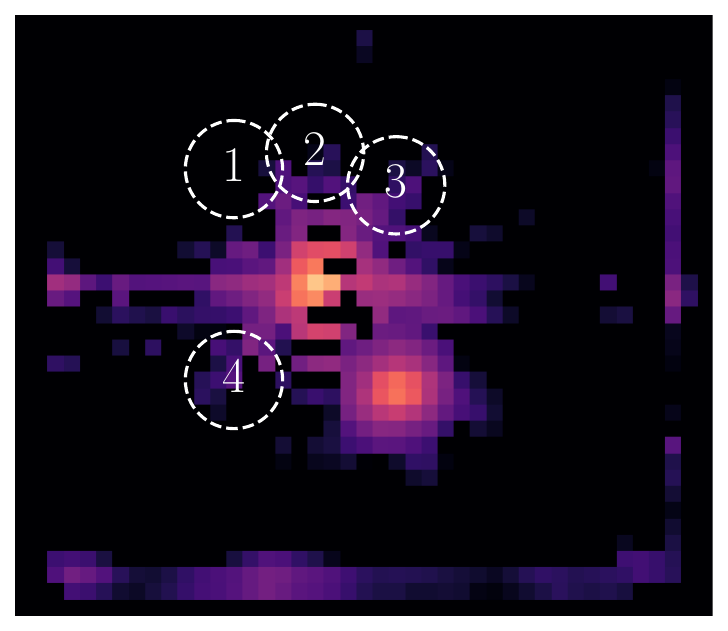} \\
\caption{The IFU image of the 2M1207 data after PSF subtraction, integrated across all wavelength elements. The dashed white circles show the positions and radii of the four areas used to sample the background flux. The orientation and color scale of the image are identical to those of Figure~\ref{fig:primary_subtraction}.}
\label{fig:2M1207_background_samples}
\end{center}
\end{figure}

As noted in the 2M1207~A subtraction step, some residuals remain in the images that may affect the measurement of 2M1207~b's flux. To investigate the potential for remaining noise in the post-subtraction spectra, we sample four regions, each centered at the same separation as 2M1207~b's centroid is from the centroid of 2M1207~A, and each with a radius of 3 pixels. These regions were then run through the same reduction pipeline as that used for 2M1207~b, yielding what should be four independent samples of background flux in time across the spectrum (Figure~\ref{fig:2M1207_background_samples}). We estimate a ``pessimistic'' noise level by calculating the mean flux at all sampled background locations, and using those values as the flux uncertainties in 2M1207~b's data where they exceed the existing estimates. These pessimistic estimates are incorporated into our estimates of the time-to-time signal-to-noise ratios (SNRs). The fluxes are small compared with the estimates of the uncertainties in 2M1207~A's spectra; therefore, we only consider their contribution to 2M1207~b's spectra. The time variability of these background samples, as well as their potential for contamination of the companion's signal, are introduced and discussed further in \S \ref{sec:variability-models:b:error-estimates}.

%%%%%%%%%%%%%%%%%%%%%%%%%%%%%%%%%%%%%%%%%%%%%%%
\section{Empirical Characterization of the Variability}
\label{sec:variability-models}
%%%%%%%%%%%%%%%%%%%%%%%%%%%%%%%%%%%%%%%%%%%%%%%
In the following subsections we will examine the spectral data as a whole, then focus on the variations in those data in time. For the latter, we define the variability ``displacement'' $S$ as the fractional difference of the spectral flux density at each time $F_\lambda\!\left(\lambda, t\right)$ from the time-mean value $\bar{F}_\lambda\!\left(\lambda\right)$:
\begin{equation}\label{eq:variability_displacement}
S\!\left(\lambda, t\right) \equiv \frac{F_\lambda - \bar{F}_\lambda}{\bar{F}_\lambda}.
\end{equation}
We calculate the SNR of this displacement across the time series following the $\sigma_\mathrm{pt}$ definition in \S 3.1 of \citet{Radigan2014a}, which is designed to be an estimate of the high-frequency noise in the light curve. This is akin to using the first order of the Taylor series to estimate the uncertainties. We calculate the standard deviation of the difference in flux between successive elements, then dividing by $\sqrt{2}$ to recover the uncertainty at each individual time. Since the light curves are not periodic, we repeat the change from the previous time element for the final time element.

\subsection{Model Definitions and Selection Criteria}\label{sec:variability-models:model_definitions}
We begin our characterization by modeling the time variation for each wavelength element with a series of functions, starting with a constant function in time and increase in polynomial order through a cubic function. The final model considered is a sinusoid in time. The nominal assumption is that, if variability is primarily driven by rotation, one expects a periodic signal that may be approximated with a sinusoid \citep[see e.g.][]{Zhou2016,Biller2017}. This is borne out, for example, in observations of the WISE1049~AB brown dwarf binary --- observations that span $\approx 100$ rotations and show a robust convergence of sinusoidal models to the rotation period \citep[][]{Apai2021,Fuda2024a}. However, were the period to be well in excess of the observing duration, we still would like to have a low-order polynomial approximation of the trend, as is done to linear order in studies such as \citet{Buenzli2014}. A lack of evidence for a sinusoidal model therefore should be interpreted as a lack of sufficient data for a robust detection of a sinusoidal signal, rather than ruling out that one exists at any period whatsoever.\footnote{The detectability of rotationally-modulated variability also intrinsically depends on the viewing angle. For example, objects viewed close to pole-on will show little to no variability from rotation \citep[see e.g.][]{Fuda2024b}.}

 The fits to all polynomial models were performed using the \texttt{polyfit} function in \texttt{numpy} \citep{van11}, with the fit weighted by the inverse of the uncertainties in the data. The model parameter uncertainties are calculated using the covariance matrix of the fit. The final model considered is a sinusoid
\begin{equation}\label{eq:sinusoidal_model}
S\!\left(\lambda, t\right) = S\!\left(\lambda, t=t_0\right) + \Delta S \cos\!\left[ \Omega t + \phi\!\left(\lambda, t=t_0\right) \right]
\end{equation}
with four free parameters: $S\!\left(\lambda, t=t_0\right)$ is the offset of the zero point of the sinusoid relative to the time-mean flux, $\Delta S$ the amplitude, $\Omega$ the frequency of the variability in time, and $\phi\!\left(\lambda, t=t_0\right)$ the phase angle. To perform the fit to this model, we use the \texttt{curve\_fit} routine in \texttt{scipy} \citep{jon01}. The fitting is weighted by the inverse of the uncertainties of the data, and the uncertainties are calculated using the covariance matrix, as with the polynomial fits. Note that the number of free parameters in the cubic model matches that of the sinusoid --- the choice to consider the latter the ``most complex'' is arbitrary but is meant to distinguish a potential detection of rotationally-modulated variability. The model comparison is done by comparing Bayesian Information Criterion (BIC) values as an approximation of the Bayes factors, with a Gaussian likelihood function that assumes no correlation in noise with wavelength or time:
\begin{align}
\label{eq:BIC}
\begin{split}
    \Delta \mathrm{BIC} =&\, n_\mathrm{par,\,simp} \log n_\mathrm{data} - 2 \log \mathcal{L}_\mathrm{simp} \\
    &- \left( n_\mathrm{par,\,comp} \log n_\mathrm{data} - 2 \log \mathcal{L}_\mathrm{comp} \right) \\
    =&\, \left(n_\mathrm{par,\,simp} - n_\mathrm{par,\,comp}\right)\log n_\mathrm{data} \\
    &- 2 \left( \log \mathcal{L}_\mathrm{simp} - \log \mathcal{L}_\mathrm{comp} \right)
\end{split}
\end{align}
where $n_\mathrm{par}$ is the number of parameters, $\log \mathcal{L}$ denotes the base-ten logarithmic likelihood (hereafter ``log likelihood'') value of each model, and the abbreviated subscripts denote the more complex (``comp'') and more simple (``simp'') model. Note we are ordering the models here such that a higher BIC denotes a preference for the more complex model. For the baseline test (line versus constant), $n_\mathrm{par, constant} = 1$, $n_\mathrm{par, line} = 2$, and $n_\mathrm{data} = 20$ is the number of samples in time. We employ our statistical tests in Equation \ref{eq:BIC} to evaluate the relative evidence for each pair of simpler versus more complex model. We adopt a conventional heuristic as a guideline \citep[see for example the interpretation of Bayes factors and BIC as its approximation in][]{Kass1995}: wavelengths where the more complex model's BIC is greater by at least 2 are considered to show a ``marginal'' positive preference; a difference of 6 shows a ``moderate'' preference; and a difference of at least 10 is considered a ``strong'' preference. To determine the ``most preferred'' model, we pick one of these thresholds for model preferences, then compare each more complex model with all models simpler than it. If a model reaches the threshold, it becomes the new minimum complexity model with which subsequent, more complex models are compared. A constant model may be preferred in cases where the flux varies rapidly (i.e.~on the order of a single exposure) and/or irregularly. The following subsections describe the results of this characterization for each object.

\subsection{2M1207~A} 
\label{sec:variability-models:A}

\begin{figure*}
\begin{center}
\includegraphics[height=20cm]{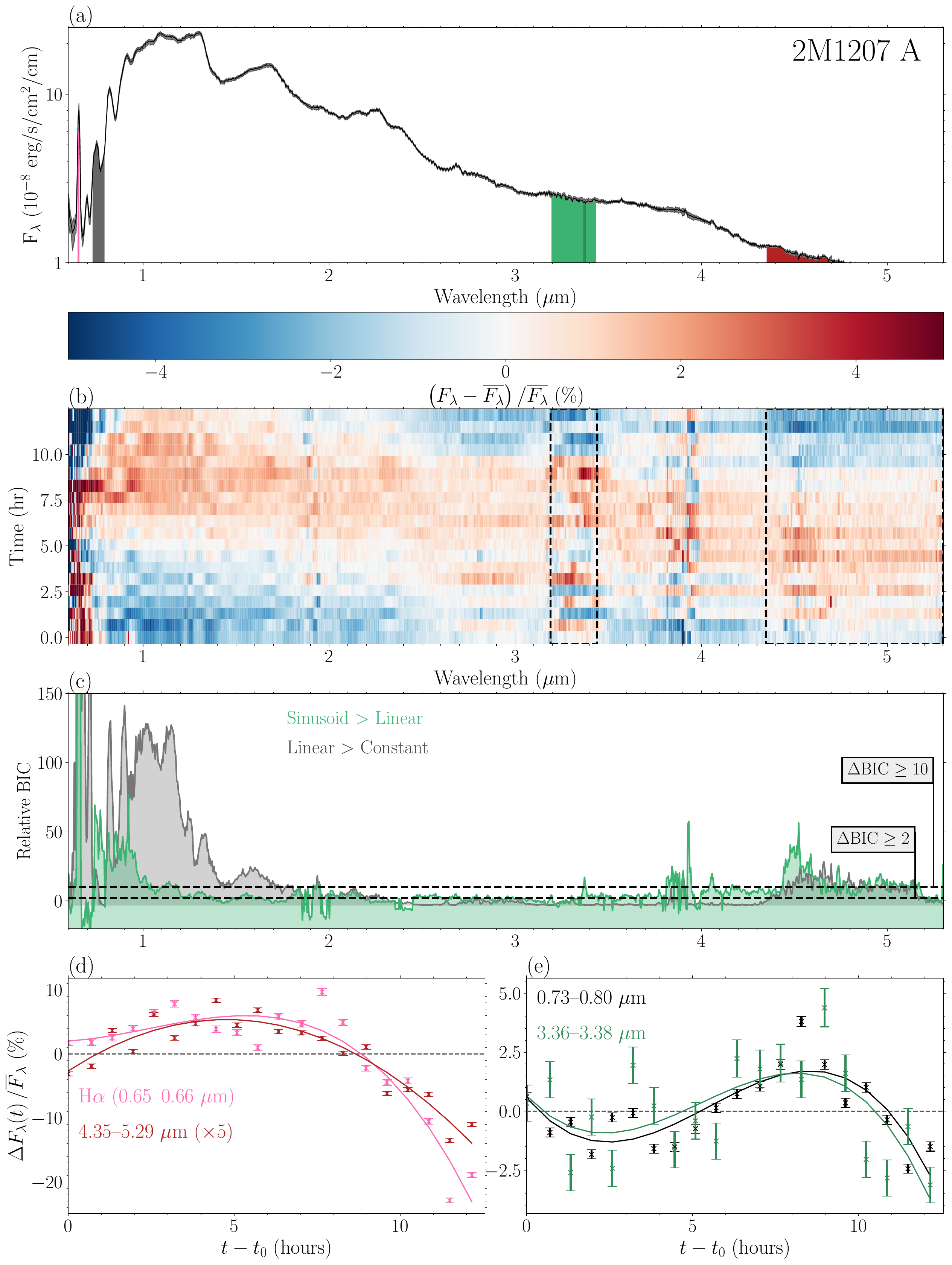}
\caption{\emph{(a):} The spectral flux density of the 2M1207~A data. The thickness of the line represents the range of values across the time series. Specific wavelength ranges are highlighted with colors that span the vertical extent below the spectrum. Each corresponds to a feature in the variability discussed in the text: H$\alpha$ in pink, 0.73--0.80~$\upmu$m in grey, 3.19--3.44~$\upmu$m in light green, 3.36--3.38~$\upmu$m in dark green, and $>4.35$~$\upmu$m in red. \emph{(b):} The variability data are shown on a color map with dimensions of wavelength versus time of observation (20 exposures across 12.56 hours). The color shows the variability displacement as defined in Equation~\ref{eq:variability_displacement}). \emph{(c):} The values of the model comparison between a linear and constant model for the variability displacement at each wavelength (gray region), considered our baseline test for a trend in time; and for a sinusoidal model versus linear at each wavelength (green region). The wavelength elements containing H$\alpha$ emission show an extreme outlier on this plot, with relative evidences for both model comparisons exceeding $\Delta$BIC $\gtrsim$ 1000. \emph{(d) and (e):} Pairs of light curves chosen based on similarities in their best-fit sinusoidal parameters. (d) shows the H$\alpha$ light curve plotted along with the light curve integrated across all wavelengths beyond 4.35~$\upmu$m. (e) shows light curves integrated across the ranges 0.73--0.80~$\upmu$m and 3.36--3.38~$\upmu$m. Each plot shows the best-fit model for each light curve; each model is the most preferred out of the choices in \S \ref{sec:variability-models:model_definitions} using a selection threshold of $\Delta$BIC $\geq 6$.}
\label{fig:variability-map-2M1207A}
\end{center}
\end{figure*}

We start by identifying the spectral features in the spectrum of 2M1207~A (Figure~\ref{fig:variability-map-2M1207A}a). We detect H$\alpha$ emission with a time-averaged SNR of $\sim 150$. We also see a trough at approximately 0.77~$\upmu$m shaped by the K\,\textsc{i} doublet (0.766 and 0.770~$\upmu$m), and another trough at 0.85~$\upmu$m due to broad absorption from VO. Although the $J$ band peak does not show any resolved atomic or molecular features, the dip centered at 1.2~$\upmu$m is shaped by the composite effect of absorption from H$_2$O, FeH, and multiple K\,\textsc{i} doublets \citep[see e.g.][]{Kirkpatrick1993}. H$_2$O also shapes the valley between the $J$ and $H$ peaks, the latter of which shows the triangular shape of a young brown dwarf with low surface gravity \citep[$\log g = 4.0\pm0.5$, as constrained from a fit to a spectral model grid in][]{Manjavacas2024}.

The amplitude of the variability is typically of order a few \% across the spectrum. The wavelength- and time-mean SNR of the 2M1207~A is $\sim 100$, which serves as a reference for the noise level relative to the typical time-to-time variation in the flux. The SNR of the displacement of $\approx 7.6$, with a range of $\approx 3$--18. To examine the structure of this variability, we show a variability map in Figure~\ref{fig:variability-map-2M1207A}b, with color representing the variability displacement across wavelength and time. The map data suggest trends in time across most wavelengths. On the broadest wavelength scale, we see a nearly contiguous band of fluxes above the time average, as shown in the red hues on the map. Moving along this band in wavelength, we see that the maximum flux tends to occur earlier in the observation. This contiguous feature is interrupted by a distinct variability feature spanning approximately 3.19--3.44~$\upmu$m that fluctuates more rapidly, on the order of a single integration time. The highest amplitude of variability by far is seen at the H$\alpha$ emission feature, with a peak-to-peak range exceeding 30\% across the time series (Figure~\ref{fig:variability-map-2M1207A}d).

\begin{figure}
\begin{center}
\includegraphics[width=8.5cm]{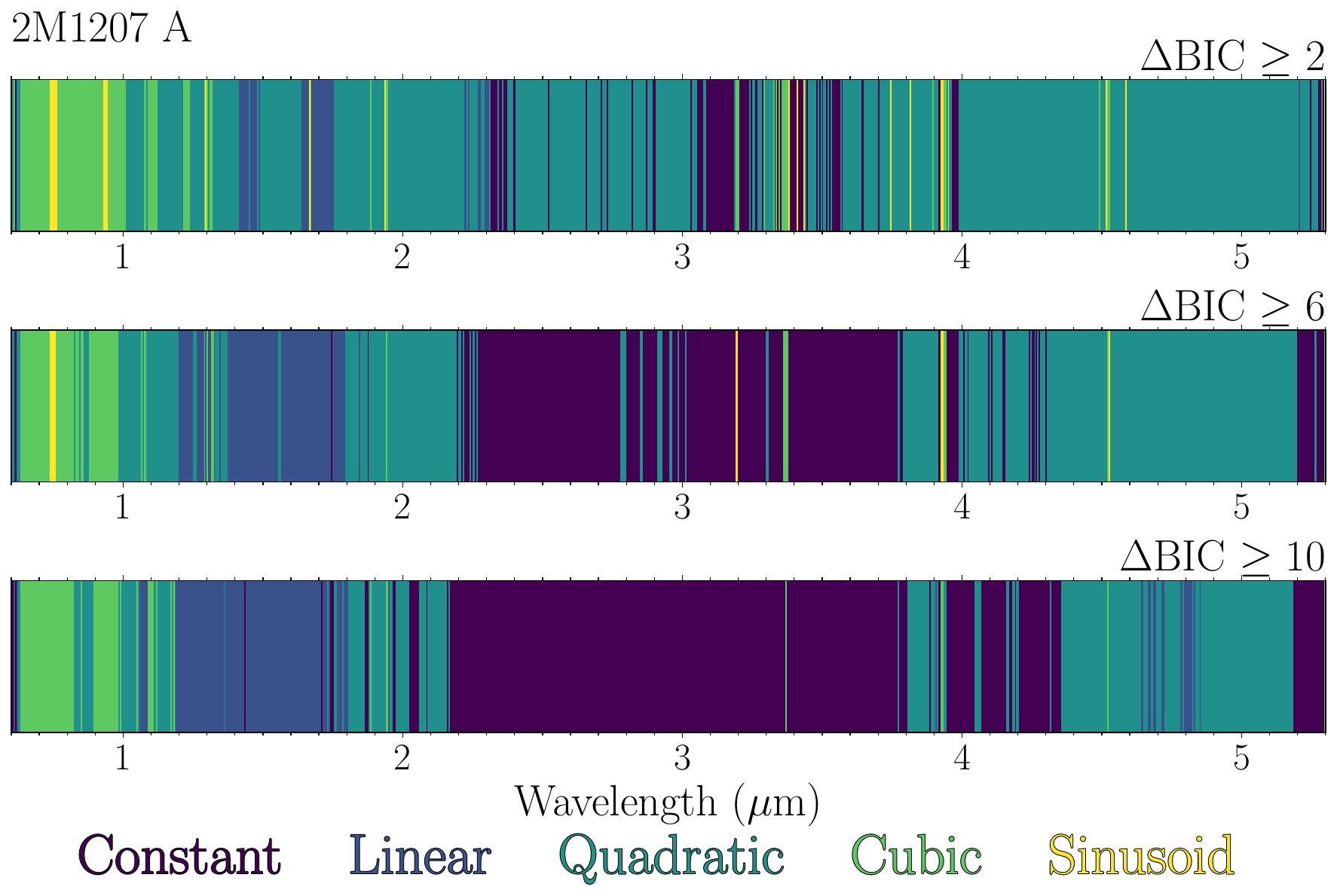} \\
\caption{Colored bars representing the preferences for models of 2M1207~A, including models of intermediate complexities (i.e.~quadratic and cubic polynomials). Each model is fit to the mean-subtracted time series data at each wavelength element. Our criteria for determining the most-preferred model is described in \S \ref{sec:variability-models:model_definitions}. The brighter the color, the more complex the model preferred at that individual wavelength. We show three thresholds for model selection: $\Delta$BIC $\geq 2$, 6, and 10.}
\label{fig:2M1207A_variability_model_BIC_comparisons}
\end{center}
\end{figure}

We select the results of two model comparisons to show in Figure~\ref{fig:variability-map-2M1207A}c: the relative evidence for the simplest trend versus no trend at all (i.e.~linear versus constant models), and for the simplest periodic trend versus the simplest trend (i.e.~sinusoidal versus linear models). There is at least marginal support ($\Delta$BIC $\geq 2$) across most of the spectrum for periodic models when compared with linear models. The results of comparisons among all models are summarized in the three color bars of Figure~\ref{fig:2M1207A_variability_model_BIC_comparisons}. This allows us to show trends that are intermediate in complexity; in other words, where the trends may be sufficiently non-linear, but do not show enough of a periodic signal to prefer the sinusoidal model. Non-constant models reach the strictest threshold ($\Delta$BIC $\geq 10$) for a large fraction of wavelengths below about 2.2~$\upmu$m and above about 3.8~$\upmu$m. Between 2.2--3.8~$\upmu$m, there is marginal preference ($\Delta$BIC $\geq 2$) for most of the wavelengths for at least a quadratic model --- cubic fits are preferred intermittently within a range spanning approximately 3.19--3.44 $\upmu$m. While only a few individual wavelengths prefer a sinusoid above all other models tested, by applying the sinusoid model uniformly across the entire spectrum, we can interpret the data in terms of the potential for signals from periodic sources of variability.

\begin{figure*}
\begin{center}
\includegraphics[height=22cm]{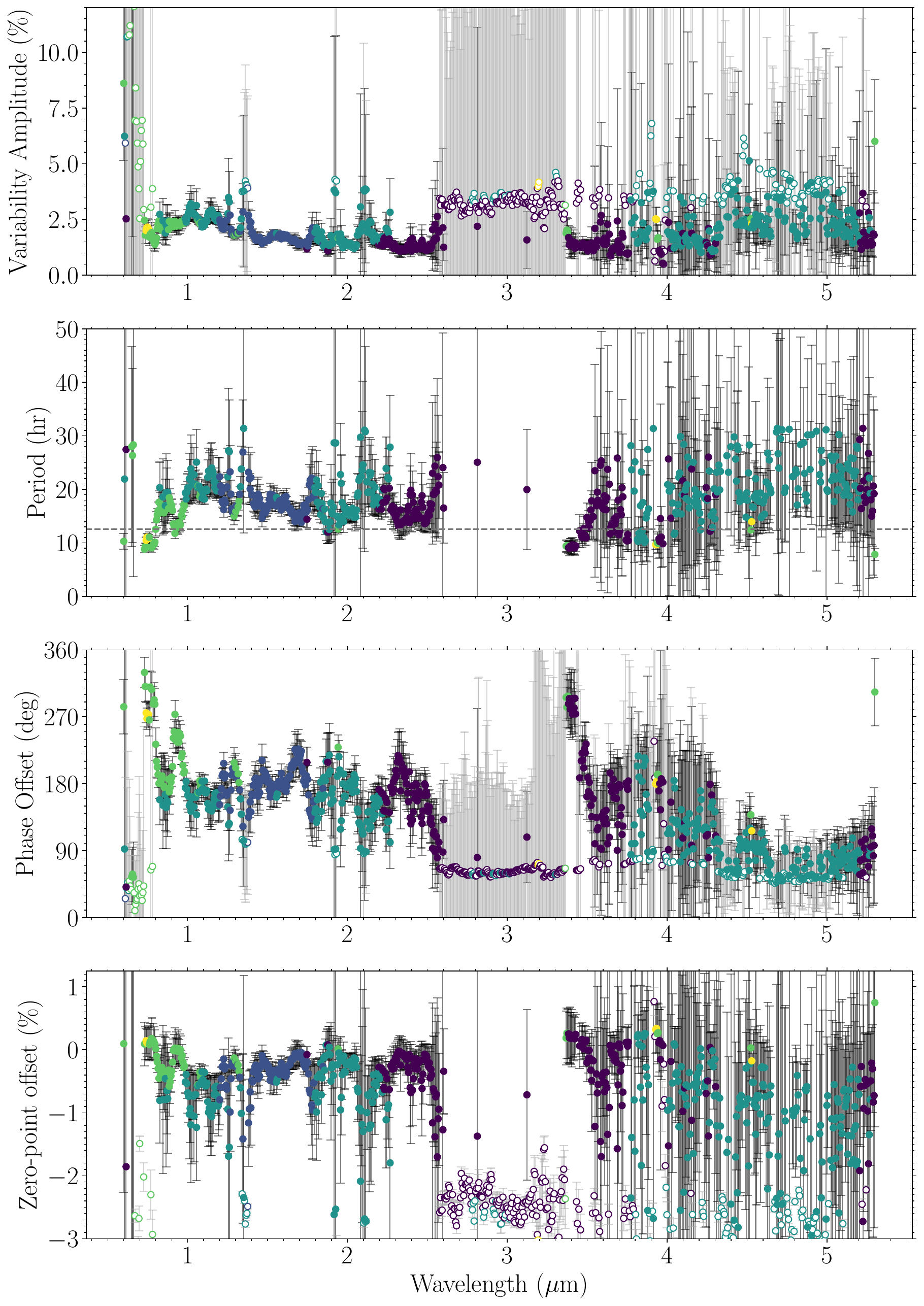}
\caption{Best-fitting values for a fit of the displacement of 2M1207~A's flux from the time-mean to a sinusoid (Equation~\ref{eq:sinusoidal_model}), as colored points with black error bars. The color of each point corresponds to the ``most preferred'' model for the ``moderate'' threshold of $\Delta$BIC $\geq 6$, matching the colors in Figure~\ref{fig:2M1207A_variability_model_BIC_comparisons}: purple is a constant model, blue a linear model, teal a quadratic model, green a cubic model, and yellow a sinusoidal model. Values with a $\Delta\mathrm{BIC}>2$ preferring a sinusoid over a linear fit are shown with filled circles; others are shown with empty circles. The plot of best-fit variability periods excludes points that return a period at the upper bound of our fit, which is $2.5\times$ the observing duration of 12.56 hours. The observing duration is shown as a dashed grey line.}
\label{fig:variability-model-2M1207A}
\end{center}
\end{figure*}

We show the results of sinusoid fits to all wavelengths in Figure~\ref{fig:variability-model-2M1207A}. Each set of best-fit parameter values shows variation across the spectrum, and many constraints have large uncertainties. With few exceptions, the best-fit variability periods exceed the observing baseline. Another feature to note is in the phases, which represent when each best-fit sinusoid reaches its maximum flux. The plot of these values suggests a trend toward later peaks with decreasing wavelength, with the fits at wavelengths beyond 4.35~$\upmu$m showing peaks that occur the earliest in the observations. We also use these sinusoidal fits to identify any isolated wavelength ranges whose light curves may suggest mutual correlations in time. As one example, the 3 wavelength elements sensitive to H$\alpha$ (0.65--0.66~$\upmu$m) have phases consistent with the data beyond 4.35~$\upmu$m. We compare the integrated light curves for these two regions in Figure~\ref{fig:variability-model-2M1207A}d. Another pair of contiguous wavelength regions, at 0.73--0.80 $\upmu$m and 3.36--3.38 $\upmu$m, both show moderate evidence for at least a cubic model and return fit parameters consistent with each other --- we show their light curves in Figure~\ref{fig:variability-model-2M1207A}e. Table \ref{table:2M1207A_lightcurve_models} shows the model fits to the light curves integrated over these wavelength ranges. Cubic models are statistically preferred for the pair of H$\alpha$ and $>4.35$~$\upmu$ light curves. The correlation is not immediately obvious from the fit coefficients, particularly because the amplitudes of their variability are different. We examine their potential correlation further in \S~\ref{sec:discussion:A}. Both the best-fit cubic and sinusoidal models for the pair of 0.73--0.80 and 3.36--3.38~$\upmu$m light curves show agreement across all parameters. We comment on the possible physical interpretations of this correlation as well in \S~\ref{sec:discussion:A}.

\begin{deluxetable}{rcccc}[htb!]\label{table:2M1207A_lightcurve_models}
\tabletypesize{\scriptsize}
\tablewidth{0pt}
\tablecaption{Fits to cubic and sinusoid models for the light curves presented in Figures \ref{fig:variability-map-2M1207A}d and e. Bolded names indicate the model most preferred by the statistical evaluation described in \ref{sec:variability-models:model_definitions}. The retrieved sinusoidal periods for the H$\alpha$ and 4.35--5.29 $\mu$m exceed the observing window; their fits are not included.}
\tablehead{
    \colhead{Wave. ($\upmu$m)} &
    \multicolumn{4}{c}{Parameter Values}
    }
\startdata
\colhead{Cubic} &
    \colhead{$c_0$ (\%)} &
    \colhead{$c_1$ (\%)} &
    \colhead{$c_2$ (\%)} &
    \colhead{$c_3$ (\%)} \\
\hline
\textbf{H$\alpha$} & $2.0\pm2.4$ & $0.3\pm1.8$ & $0.3\pm0.4$ & $-0.04\pm0.02$\\
\textbf{4.35--5.29} & $0.7\pm0.3$ & $0.9\pm0.2$ & $-0.12\pm0.04$ & $0.003\pm0.002$\\
0.73--0.80 & $0.6\pm0.7$ & $-1.7\pm0.5$ & $0.4\pm0.1$ & $-0.027\pm0.005$\\
\textbf{3.36--3.38} & $0.6\pm1.3$ & $-1.5\pm0.9$ & $0.4\pm0.2$ & $-0.03\pm0.01$\\
\vspace{0.1cm} \\
\colhead{Sinusoid} &
    \colhead{$\Delta S$ (\%)} &
    \colhead{$P_\mathrm{var}$ (hr)} &
    \colhead{$\phi\!\left(\lambda, t=t_0\right)$ (${}^\circ$)} &
    \colhead{$S\!\left(\lambda, t=t_0\right)$ (\%)} \\
\hline
\textbf{0.73--0.80} & $1.8\pm0.3$ & $9.8\pm0.6$ & $291\pm16$ & $0.2\pm0.2$\\
3.36--3.38 & $1.9\pm0.6$ & $9.3\pm1.1$ & $294\pm30$ & $0.2\pm0.4$\\
\enddata
\end{deluxetable}

\subsection{2M1207~b}
\label{sec:variability-models:b}

\subsubsection{Accounting for Flux Uncertainties}\label{sec:variability-models:b:error-estimates}
\begin{figure*}[htb!]
\begin{center}
\begin{tabular}{cc}
\multicolumn{2}{c}{\includegraphics[width=17cm]{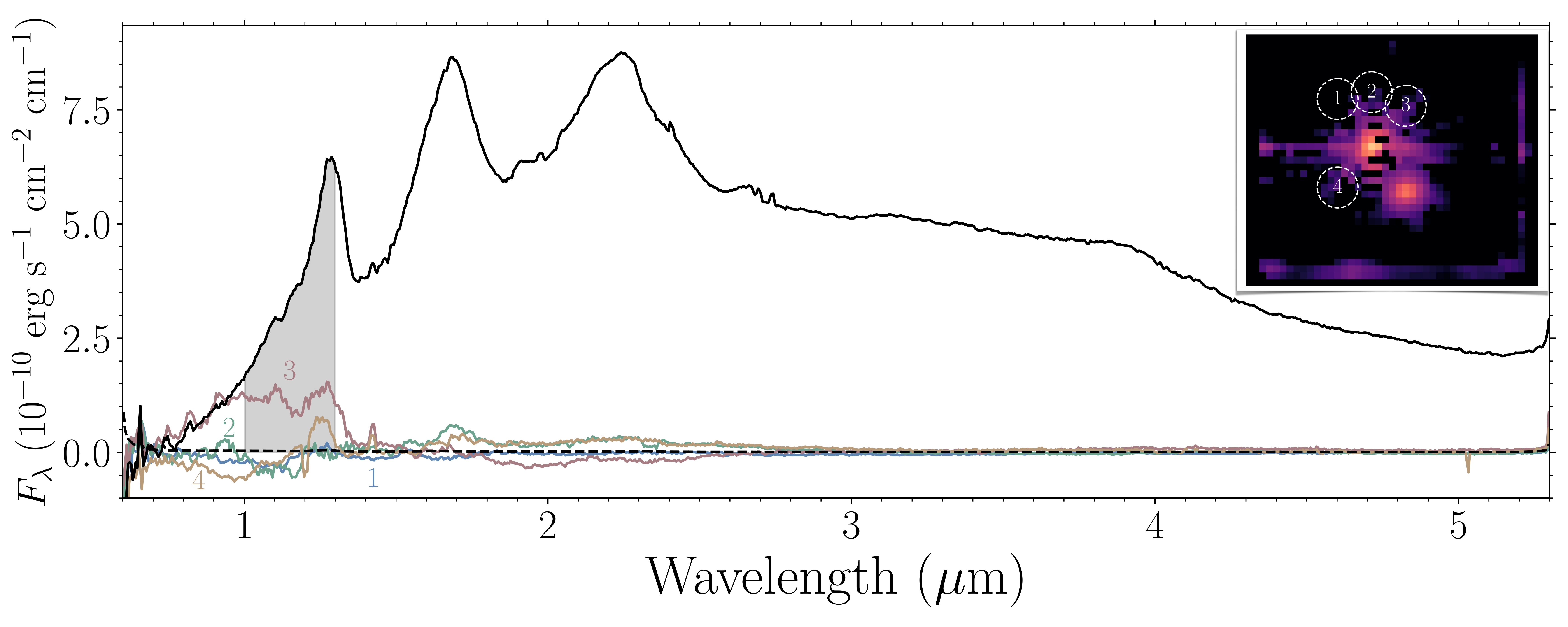}} \\
\includegraphics[height=8cm]{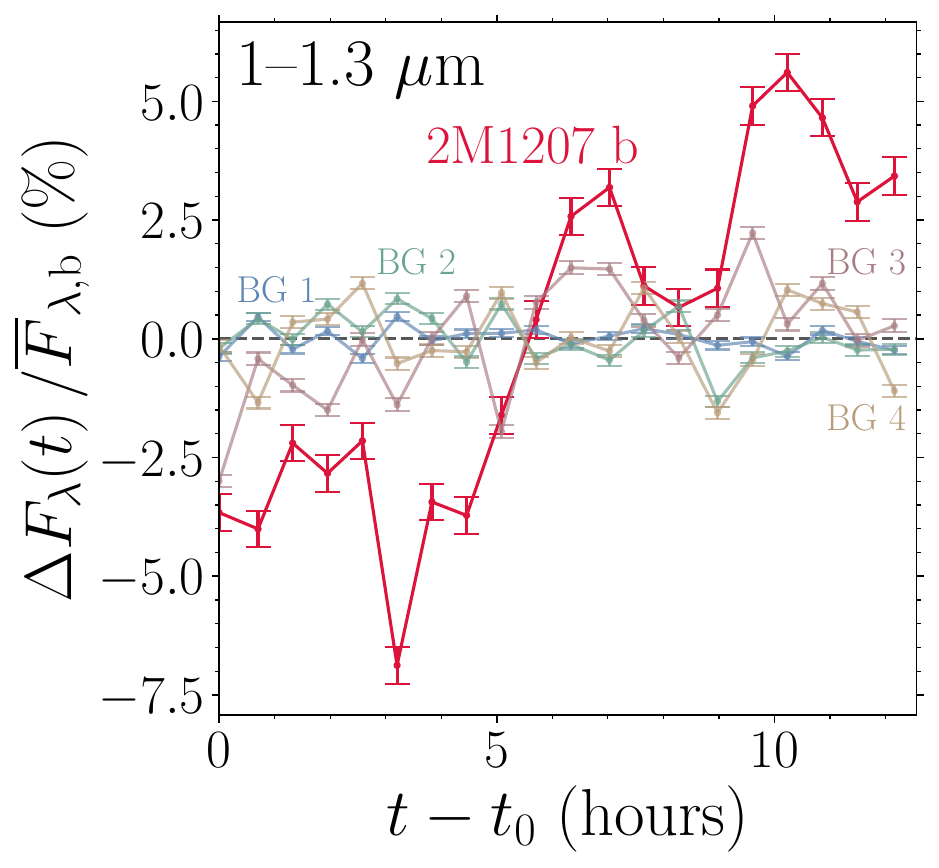} &
\includegraphics[height=8cm]{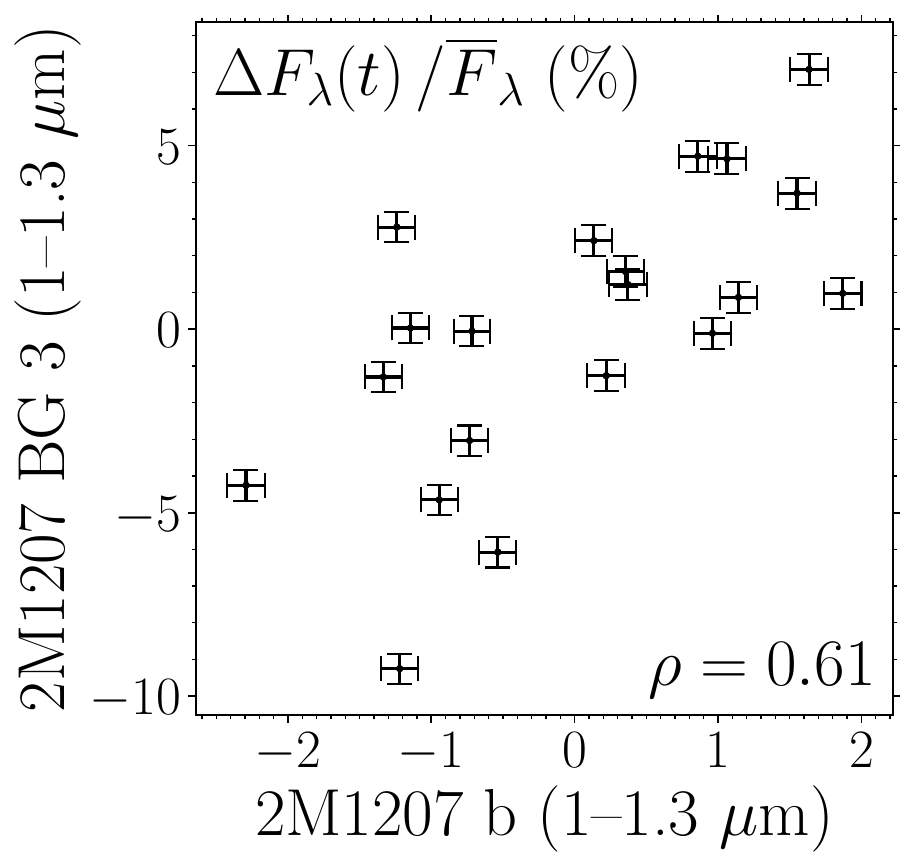}
\end{tabular}
\caption{A reproduction of the time-mean 2M1207~b spectrum with the corresponding fluxes measured by processing each of the background samples in the same way as was done to reduce the 2M1207~b data. The grey region shows the range 1--1.3~$\upmu$m, an approximation of where any of the measured background fluxes is $\gtrsim$10\% of the flux of 2M1207~b. \emph{Top inset:} A reproduction of Figure \ref{fig:2M1207_background_samples}, for reference. \emph{Bottom left:} A comparison of the integrated light curve of the variability displacement of 2M1207~b from 1--1.3~$\upmu$m, with a light curve of the background samples integrated over the same wavelengths, normalized to the time-mean flux of 2M1207~b. \emph{Bottom right:} The variability displacements of each light curve, here normalized to their respective time-means. $\rho$ represents the coefficient from a Spearman rank correlation test; here, we see the light curves are moderately correlated at $\rho=0.61$. }
\label{fig:2M1207b_vs_background}
\end{center}
\end{figure*}

As introduced in \ref{sec:observations:post_subtraction_residuals}, we use samples of background regions in the images of the 2M1207 system to account for any residual uncertainties in the spectra of 2M1207~b. The spectra of these background regions are compared with that of 2M1207~b in Figure~\ref{fig:2M1207b_vs_background}. We identify one of the regions, shown in Figure~\ref{fig:2M1207_background_samples} and as an inset in Figure~\ref{fig:2M1207b_vs_background} with the number 3, as having the highest average noise level across the spectrum of those sampled. We compare both the relative flux levels and the potential for correlation in time between this background sample and 2M1207~b. The background flux closely matches that of 2M1207~b until approximately 1~$\upmu$m, after which it levels out at $F_\lambda\sim10^{-10}$\,erg\,s$^{-1}$\,cm$^{-2}$\,cm$^{-1}$. This is $\sim 60$~\% of the flux of 2M1207~b, dropping to $\sim 6$~\% by 1.3~$\upmu$m. We investigate the potential for time-correlated noise by comparing the 1--1.3~$\upmu$m light curves of 2M1207~b and background 3, as shown in the bottom-left panel of Figure~\ref{fig:2M1207b_vs_background}. We find the light curves are moderately correlated (bottom-right panel of Figure~\ref{fig:2M1207b_vs_background}), with a Spearman rank correlation coefficient of $\rho=0.61$. Therefore, while the background shows a time correlation with the companion, it cannot fully account for the signal, permitting us to interpret the data in this range as a potential astrophysical signal.

\subsubsection{Features and Variability Characterization}\label{sec:variability-models:b:characterization}
\begin{figure*}
\begin{center}
\includegraphics[height=20cm]{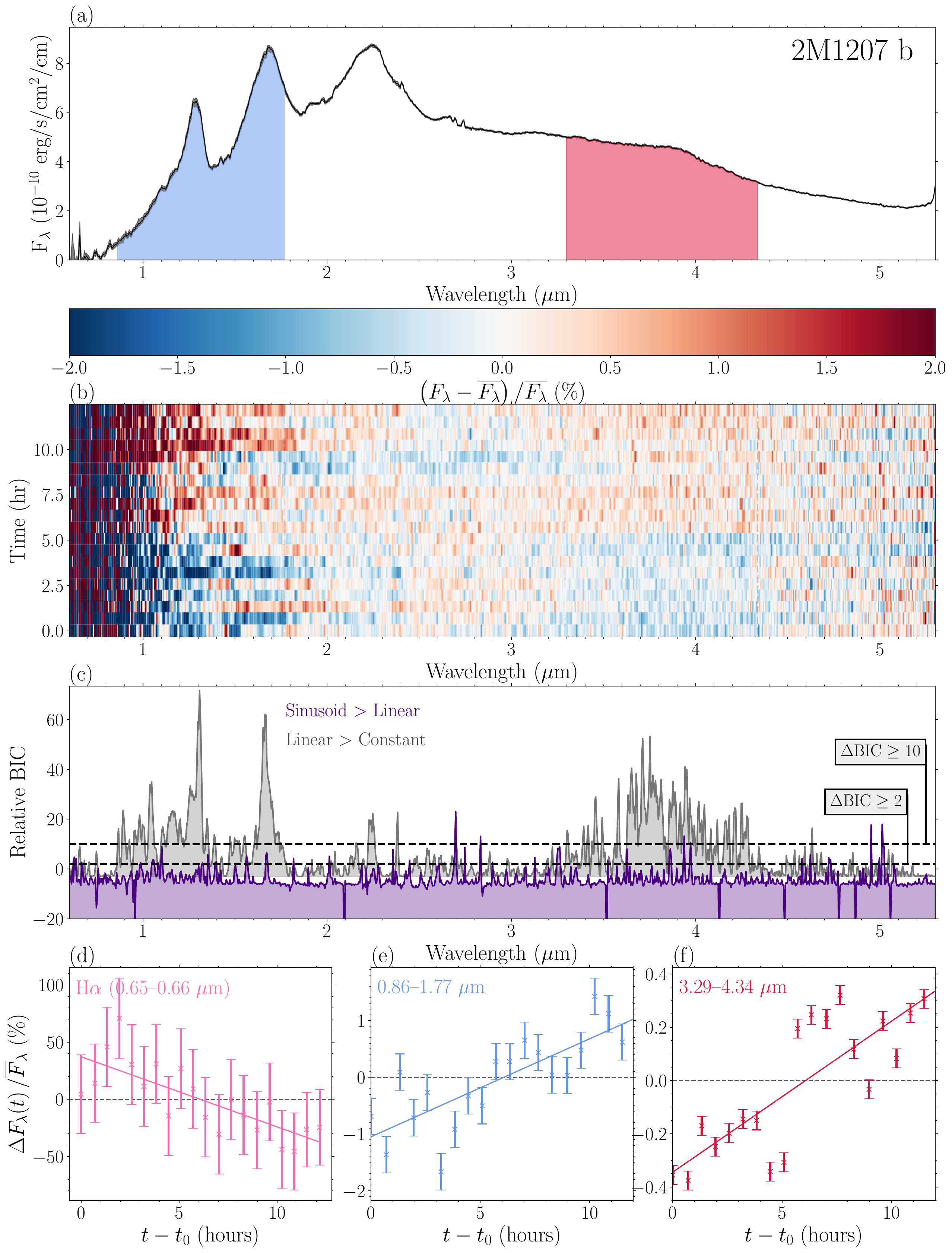}
\caption{\emph{(a):} Spectral flux density of the 2M1207~b data. The thickness of the line represents the range in time across the observation. \emph{(b):} The displacements of the flux from the time-mean (Equation~\ref{eq:variability_displacement}), shown on a color map with dimensions of wavelength versus time of observation (20 exposures across 12.56 hours). \emph{(c):} The relative evidence for a linear versus flat model in variability displacement (via Equation \ref{eq:BIC}, shown as the gray region), and for a sinusoidal model versus linear at each wavelength (purple region). The two wavelength-integrated light curves with the most evidence for non-constant trends are 0.86--1.77~$\upmu$m and 3.29--4.34~$\upmu$m. \emph{(d)--(f):} Light curves showing the variability displacement for H$\alpha$ (d) and across the two wavelength ranges with the highest contiguous evidence for a linear trend in variability displacement across time (e) and (f). All light curves return the strongest preferences for linear models which are plotted over the data.}
\label{fig:variability-map-2M1207b}
\end{center}
\end{figure*}

The PRISM spectra of 2M1207~b (Figure~\ref{fig:variability-map-2M1207b}a) shows broad absorption from H$_2$O across the spectrum, shaping the $J$, $H$, and $K$-band peaks. The CO bands starting at approximately 2.3~$\upmu$m shape the red half of the $K$-band peak, but no individual line is resolved besides H$\alpha$ emission. The time-averaged SNR of the H$\alpha$ emission feature is $\approx 7$. The SNR for each spectrum in time remains of order unity until approximately 0.8~$\upmu$m, beyond which it improves rapidly. The time- and wavelength-mean SNR across the entire spectrum $\sim$300 with the original uncertainty estimates, dropping to $\sim$185 if the pessimistic noise profile is adopted. The 2M1207~b data has a higher SNR relative to 2M1207~A because there is saturation in 2M1207~A's central pixels which introduces additional model uncertainty in the multi-stage PSF fit.

\begin{figure}
\begin{center}
\includegraphics[width=8.5cm]{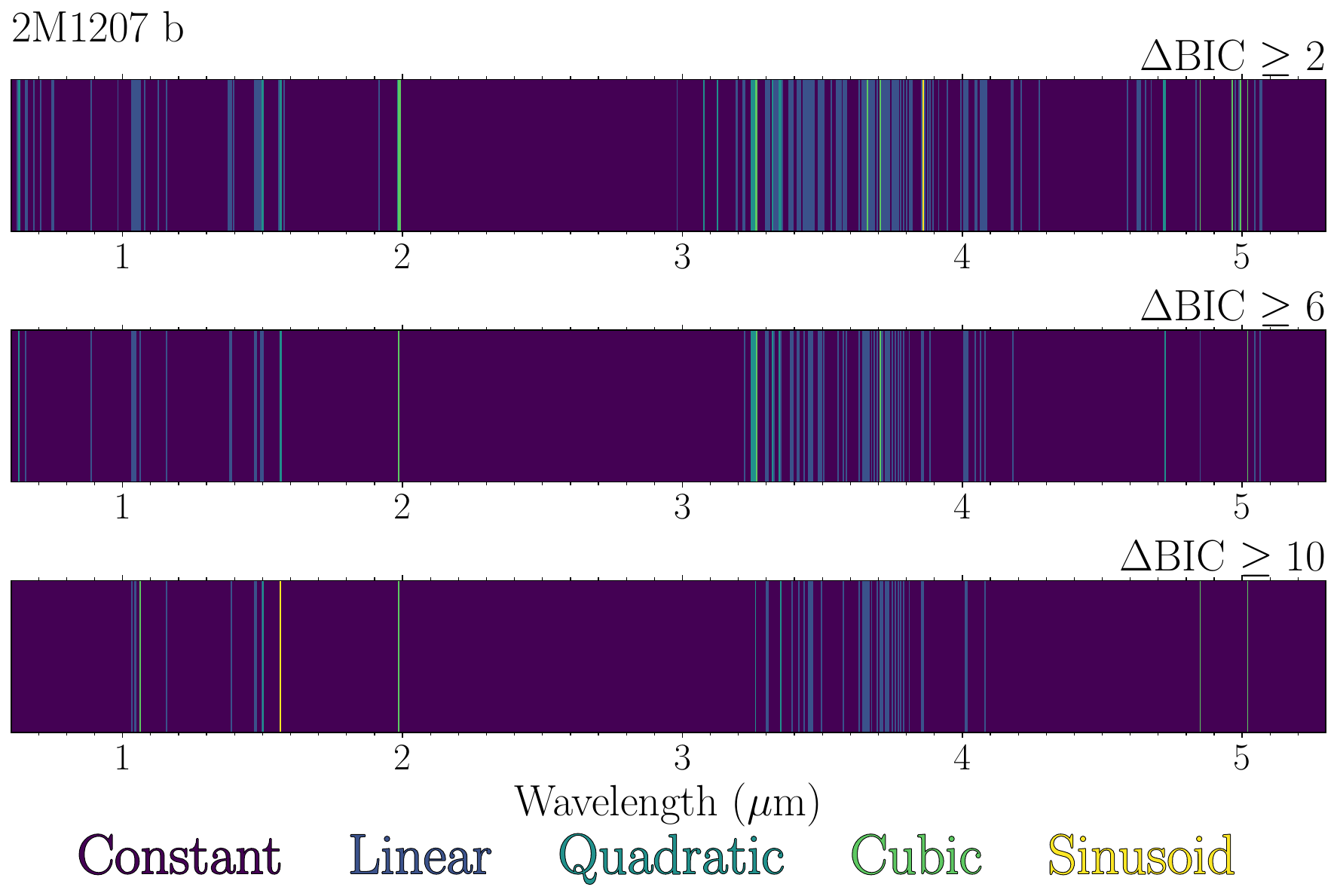} \\
\caption{Colored bars representing the preferences for models of 2M1207~b, including models of intermediate complexities (i.e.~quadratic and cubic polynomials). Each model is fit to the mean-subtracted time series data at each wavelength element. Our criteria for determining the most-preferred model is described in \S \ref{sec:variability-models:model_definitions}. The brighter the color, the more complex the model preferred at that individual wavelength. We show three thresholds for model selection: $\Delta$BIC $\geq 2$, 6, and 10.}
\label{fig:2M1207b_variability_model_BIC_comparisons}
\end{center}
\end{figure}

The trends of 2M1207~b's variability (Figure~\ref{fig:variability-map-2M1207b}b) are less obvious by eye than for 2M1207~A. The blue end of the spectrum ($\lesssim 0.9$~$\upmu$m) is much noisier than in 2M1207~A, which limits our analysis in this range. The total displacement range in the H$\alpha$ feature, whose light curve is shown in Figure~\ref{fig:variability-map-2M1207b}d, is in excess of 100\%, with a maximum seen approximately 2.5 hours into the observation and a minimum at 10 hours. Broadening our inspection, there are areas in the spectrum that suggest an increase in flux across the observation. To quantify these potential trends, we calculate the relative evidences for models of complexities between constant and sinusoid; the BIC comparisons for linear versus constant and sinusoid versus linear are highlighted in Figure~\ref{fig:variability-map-2M1207b}c, with model preferences across all models illustrated in Figure~\ref{fig:2M1207b_variability_model_BIC_comparisons}. There are regions that have intermittent support for at least a linear model, but there is little support for more complex models. Since the signal per wavelength element is low, we choose to integrate across wavelength to maximize what evidence does exist for a potential trend in time. By initial inspection, the strongest evidence for linear models appears to come in two peaks that align with the peaks of the $J$ and $H$ bands, at 1.29 and 1.68~$\upmu$m, respectively. We also see a broader range between approximately 3.3--4.3~$\upmu$m with strong evidence for a linear trend.

To make precise the boundaries of these two high-evidence areas in the spectrum, we define variability ``bands'' --- contiguous wavelength regions that contain the most evidence for a linear trend over no trend whatsoever. That is, were we to widen the wavelength range of these regions, their evidence would not increase. Note that even though many wavelength elements within a range may not by themselves support a linear model, they may contribute enough evidence in a wavelength-integrated light curve. We start with a light curve integrated over just two wavelength elements, at the beginning of the spectrum, and approximate the evidence using the BIC calculation in Equation \ref{eq:BIC}. We then run the statistical test on increasingly wider ranges until the evidence fails to increase. We apply this technique iteratively, placing the starting wavelength of the subset one farther along the spectrum, until we have explored all possible subsets. This returns a bound of 0.86--4.87~$\upmu$m.

Since our inspection suggested two distinct regions showing possible trends, we then divide our search range in two, testing wavelengths both smaller and larger than 2.5~$\upmu$m. The first search yields a range of 0.86--1.77~$\upmu$m, encompassing the $J$ and $H$ band peaks that show the strongest evidence. The second search bounds the region more precisely to between 3.29--4.34~$\upmu$m. The light curves integrated over each of these regions each have a $\Delta$BIC~$\sim3\times10^3$ for a linear trend over a constant model. Since the uncertainty propagation here assumes no correlation in wavelength, this represents an upper bound. The light curves are shown in Figures~\ref{fig:variability-map-2M1207b}e and~f. The shape of these two light curves is notably different, despite crossing the time-mean (zero point) concurrently. The time-mean flux at 0.86–1.77~$\upmu$m increases roughly linearly by a total of around 3\%\ over 12.56~hr, whereas at 3.29–4.34~$\upmu$m, it increases by less 1\%, also coarsely linearly. The linear fit coefficients of the wavelength-integrated light curves, as well as that of the H$\alpha$-sensitive wavelengths, are given in Table \ref{table:2M1207b_lightcurve_models}. The following section aims to contextualize these findings on 2M1207~b's variability by examining recent atmospheric modeling work based on earlier JWST data.

\begin{deluxetable}{rcc}[htb!]\label{table:2M1207b_lightcurve_models}
\tabletypesize{\normalsize}
\tablewidth{0pt}
\tablecaption{Fits to linear models for the light curves presented in Figure \ref{fig:variability-map-2M1207b}d--f.}
\tablehead{
    \colhead{Wavelengths ($\upmu$m)} &
    \multicolumn{2}{c}{Parameter Values}
    }
\startdata
\colhead{Linear Model} &
    \colhead{$c_0$ (\%)} &
    \colhead{$c_1$ (\%)} \\
\hline
H$\alpha$ & $37\pm9$ & $-6\pm1$ \\
0.86--1.77 & $-1.0\pm0.2$ & $0.17\pm0.03$ \\
3.29--4.34 & $-0.34\pm0.06$ & $0.057\pm0.009$ \\
\enddata
\end{deluxetable}

%%%%%%%%%%%%%%%%%%%%%%%%%%%%%%%%%%%%%%%%%%%%%%%
\section{Comparing the Short-Term Spectral Time Variability of 2M1207~b with Cloudy Atmospheric Models} \label{sec:atmosphere-models}
%%%%%%%%%%%%%%%%%%%%%%%%%%%%%%%%%%%%%%%%%%%%%%%

\begin{figure*}
\begin{center}
\includegraphics[width=17cm]{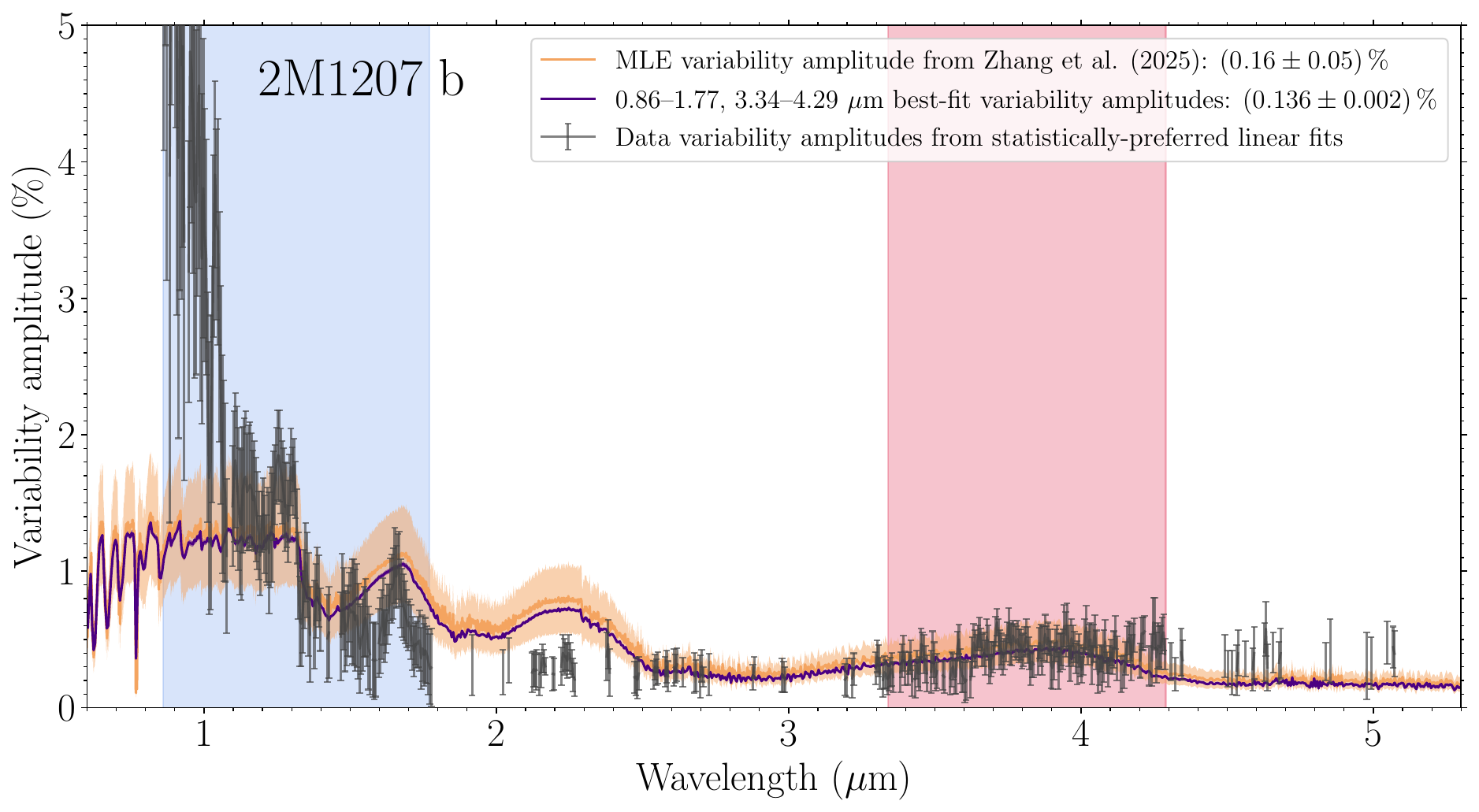}
\caption{Spectral variability amplitudes for the time-series data and a series of forward models for 2M1207~b. The individual variability amplitudes are shown in black, representing only those wavelength elements whose time series return a statistically significant linear fit versus a flat model. The orange curve and shaded region represent the maximum-likelihood and 68\%\ credible interval, respectively, of the variability amplitudes derived from varying the areal filling fraction of the two-component model presented in \citet{Zhang2025} (termed ``\texttt{QEQ-1}''). The purple curve represents the equivalent values for our reproduction of the same forward model, but where the absolute value and variation in the filling fractions are fit to our data. The model is fit only to the two regions as identified from our empirical characterization (0.86--1.77~$\upmu$m, in Figure~\ref{fig:variability-map-2M1207b}e; and 3.29--4.34~$\upmu$m, in Figure~\ref{fig:variability-map-2M1207b}f).}
\label{fig:ZJ_forward-model-spectra_variability}
\end{center}
\end{figure*}

We would like to compare our time-variable data with current atmospheric models that predict variability amplitudes as a function of wavelength. We start by comparing our data with the existing $R {\sim}2700$ NIRSpec IFU spectrum from GTO Program 1270 \citep[PI: Birkmann, as published in][]{Luhman2023,Manjavacas2024}. These data were used in a recent retrieval study \citep{Zhang2025}; therefore, these atmospheric models are useful as a starting point to characterize our new time series data. Figure \ref{fig:ZJ_forward-model-spectra_variability} shows both their data and the range in flux of our time-series spectra. We see considerable differences between these two datasets, regardless of time. The greatest differences are seen in the gap between the $H$ and $K$ band peaks, and in the thermal continuum between 3--5~$\upmu$m. This level of disagreement reaches tens of times our pessimistic measurement uncertainties --- as seen in the bottom panel of Figure \ref{fig:ZJ_forward-model-spectra_variability} --- and suggests that differences in the methods of flux calibration used between \citet{Luhman2023} and our work, and/or astrophysical variability on time scales longer than our observation, may contribute to these offsets. 

The possibility for long-term variability motivates a re-fitting of the time-variable components of the \citet{Zhang2025} atmospheric model. A key feature of the model configuration is the combination of two vertical models that represent two independent cloud thicknesses and vertical positions, which yields two distinct ``thin''-ly and ``thick''-ly clouded columns. The study also describes a model of time variability that varies the relative covering fractions of these two atmospheric columns. Their results for their most statistically-preferred model (``\texttt{QEQ-1}'', denoting Quenched Equilibrium Chemistry) are consistent with previous observations of variability in HST near-infrared bands \citep{Zhou2016}. This model configuration is outlined in the original publication, and is also summarized in \S \ref{sec:appendix:2M1207b_atmospheric_model_configuration}. By reproducing their model and associated mechanism for time variability, we can examine whether variations in cloud filling fractions are consistent not just with HST, but across the JWST NIRSpec/PRISM wavelength range.

\begin{figure*}
\begin{center}
\includegraphics[width=17cm]{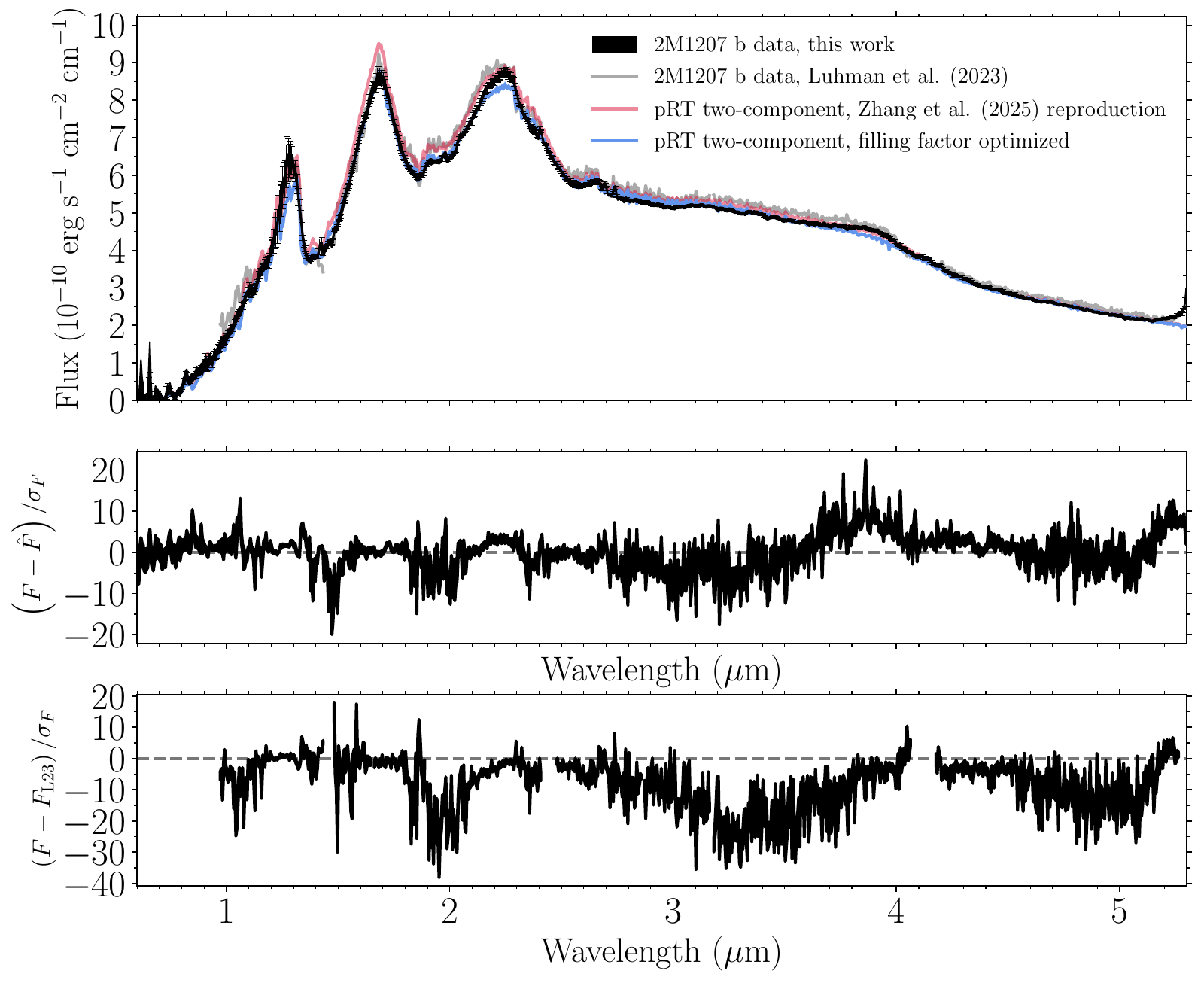}
\caption{\emph{Top panel:} Time-resolved 2M1207~b spectra (black), with the vertical thickness of the line showing the range in time. Over-plotted are error bars representing the time-mean uncertainties. The the JWST NIRSpec data published in \citet{Luhman2023} (L23) are shown in gray, and a reproduction of the ``\texttt{QEQ-1}'' model from \citet{Zhang2025} (Z25) is shown in red. A modified version of their model, with the relative areal filling fraction of the ``thin'' and ``thick'' columns optimized for our data is shown in blue. \emph{Central panel:} The residuals of each of our spectra with models optimized for cloud filling factor are shown in black, using the pessimistic data uncertainties described in \S \ref{sec:observations:post_subtraction_residuals} and \S \ref{sec:variability-models:b:error-estimates}. \emph{Bottom panel:} the range of differences across all time-resolved spectra presented in this work, each versus the down-sampled L23 spectrum, scaled to the measurement uncertainties for our spectra.}
\label{fig:2M1207b_forward-model-spectra}
\end{center}
\end{figure*}

\begin{figure}
\begin{center}
\includegraphics[width=8.5cm]{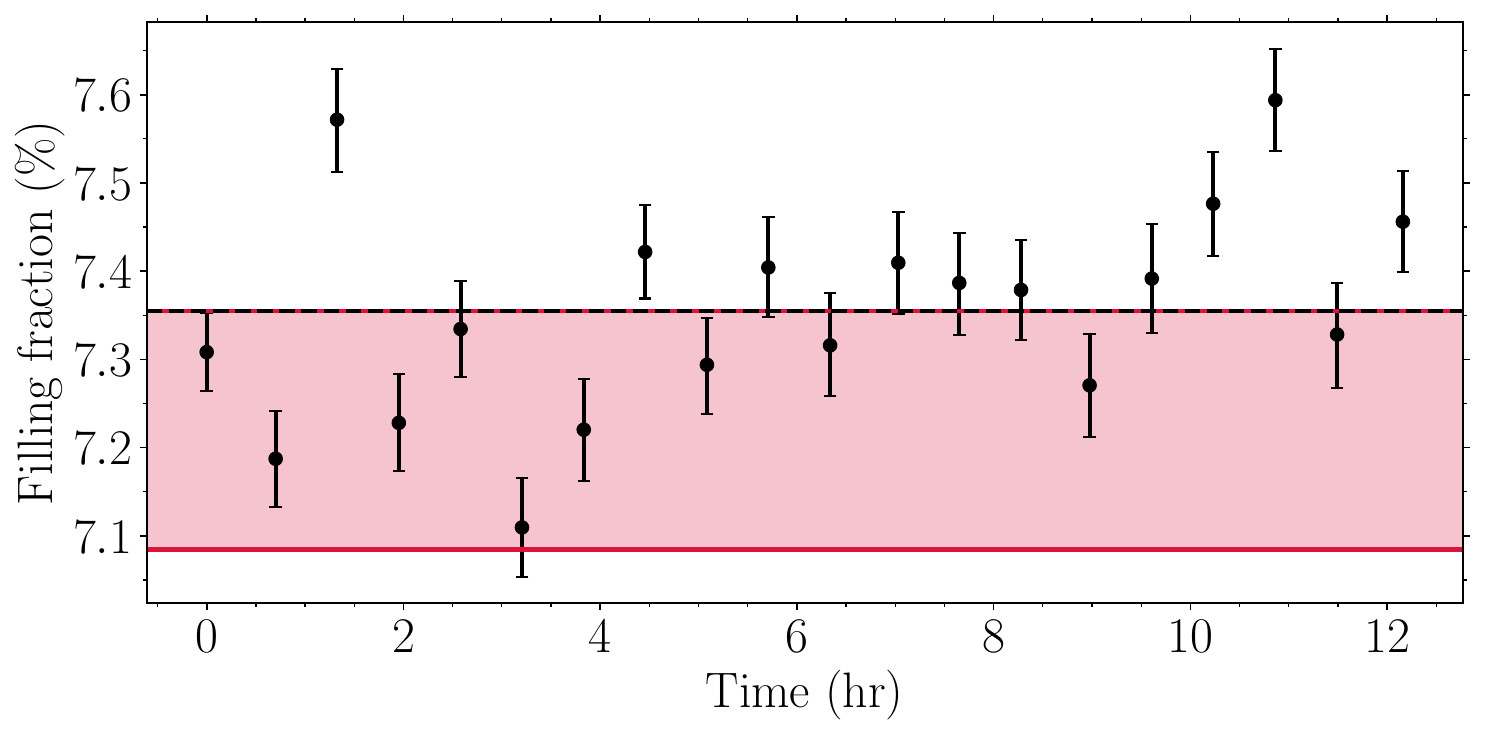}
\caption{The best-fit filling fractions of the thinly-clouded layers in the atmospheric models described in \S \ref{sec:atmosphere-models}. The dashed black line is the time-mean filling fraction. The region shaded in red represents the range of filling fractions used in the model of variability (see Figure \ref{fig:ZJ_forward-model-spectra_variability}).}
\label{fig:ZJ_refit_cloud_filling_fractions}
\end{center}
\end{figure}

We start by reproducing the maximum-likelihood forward model spectrum of \citet{Zhang2025}, using the \texttt{petitRADTRANS} \citep[abbreviated \texttt{pRT}, see][]{Molliere2020,Nasedkin2024a} code as was employed in their study. This model spectrum is shown in red in Figure \ref{fig:ZJ_forward-model-spectra_variability}. We then perform a ``mini-retrieval'' on our data at each time step, by varying only the cloud filling fractions from the previous maximum-likelihood model. This model has three parameters:
\begin{itemize}
    \item the filling fraction of the thinly-clouded layer, drawn from a uniform sample $\mathcal{U}\sim\left[0, 1\right]$;
    \item a uniform scaling term for the spectral flux, drawn from a uniform sample $\mathcal{U}\sim\left[0.5, 1.5\right]$; and
    \item a constant error scaling term, drawn from a uniform sample $\mathcal{U}\sim\left[1, 100\right]$.
\end{itemize}
This model assumes that the source of short-term variability --- on the time scale of our observing window --- is driven by clouds. The latter two parameters are included to quantify the extent of the observed differences from the previous data that cannot be captured by varying the cloud fractions alone, including potential calibration differences as well as long-term variability. The fit was performed using \texttt{NAUTILUS} \citep{Nautilus}, a general-purpose astronomical model fitting code that uses importance sampling and neural networks. The fits to our spectra are shown in blue in Figure~\ref{fig:2M1207b_forward-model-spectra}. Figure \ref{fig:ZJ_refit_cloud_filling_fractions} shows the results for the filling fractions for the thinly-clouded layer, with a range 7.12--7.59\% across the time series, and each with a model uncertainty $\approx 0.05$--0.06\%. Compared with $8.8\pm0.2$\% as constrained in the original retrieval study of \citet{Zhang2025}, our uncertainties are under-estimated. This is an expected limitation, as we cannot fully capture the marginalized posterior distribution on the filling fraction. Our error inflation factors range from 2.4--3.9, and all time steps return flux scaling factors close to $\sim 1$\% with little variation in time. The error inflation factors reflect our model residuals which reach $\sim 10$ times the flux uncertainties. In particular, the model over-predicts the shapes of the $J$ and $K$ band peaks, and under-predicts the flux both in the region between the $H$ and $K$ band peaks and in the thermal infrared continuum from $\approx3$--4~$\upmu$m.

We measure the variability amplitudes from the linear fits in \S \ref{sec:variability-models:b} within each wavelength element as one half of the slopes of the fits, multiplied by the duration of the observation. We then optimize the amount of relative change in the filling fraction of the two model atmospheric columns whose resulting changes in flux most closely match overall the observed variability amplitudes. This is done by starting with the time-mean best-fit cloud filling fraction, approximately 7.35\%, and fitting to the amount of change in filling fraction both below and above this mean value. This 2-parameter model was also fit with \texttt{NAUTILUS}, with uniform priors on the lower and upper bounds of the filling fractions spanning all possible physical values, but with no additional error inflation terms. We limit our fitted data to the union of the two wavelength ranges previously identified as containing evidence for linear trends (\S \ref{sec:variability-models:b:characterization}): 0.86--1.77 and 3.29--4.34~$\upmu$m. The best fit prefers our mean value as an upper limit, with a lower bound of approximately 7.08\%; this range is shown as the red region in Figure \ref{fig:ZJ_refit_cloud_filling_fractions}. The best-fit variability semi-amplitude accordingly is $0.136\pm0.002$\%, consistent with the $0.16\pm0.05$\% constraint from \citet{Zhang2025}. The results of this variability in wavelength space are shown in Figure~\ref{fig:ZJ_forward-model-spectra_variability}. As with the uncertainties from our fit to the filling fraction, our uncertainties here are likely to be under-estimated.

Despite broad agreement with previous results, there are three regions where the models do not fit the data well. The models fail to capture the marked rise in variability amplitudes that occur at $\lesssim 1.1$~$\upmu$m, where the models remain relatively flat. There is also an over-prediction of the amplitudes in the $K$ band. While the amplitudes beyond 4.34~$\upmu$m are consistently under-predicted by both models, the vast majority are still consistent within their uncertainties. We compare the quality of fit of the data-derived variability amplitudes versus the variability amplitudes predicted from varying the cloud filling fraction parameter in the atmospheric models in \S~\ref{sec:discussion:b}.

%%%%%%%%%%%%%%%%%%%%%%%%%%%%%%%%%%%%%%%%%%%%%%%
\section{Discussion}
\label{sec:discussion}
%%%%%%%%%%%%%%%%%%%%%%%%%%%%%%%%%%%%%%%%%%%%%%%
\subsection{Interpreting the Variability in 2M1207~A}
\label{sec:discussion:A}

\subsubsection{Trends in Periodic Phase Angle with Wavelength}\label{sec:discussion:A:periodic}
All but a handful of the variability periods derived from the sinusoidal models applied to the 2M1207~A data exceed the observing duration, which makes constraining the rotation period challenging. We expect the period constraints to be consistent with each other under a few conditions: if the variability signals at each wavelength primarily come from rotational modulation, if the observing baseline is longer than the true period, and if the evolution of the source(s) of variability do not occur on time scales shorter than the rotation period. While the period constraints are consistent for all but a couple of isolated wavelengths, the uncertainties in the periods are also similar to the observing baseline. This limits us to how precisely we can constrain a coherent periodicity. However, we do identify a trend in wavelength of the retrieved phases of our sinusoids (as shown in Figure~\ref{fig:variability-model-2M1207A}) --- broadly speaking, longer wavelengths show a peak flux earlier in the observation. We start with the phases within approximately 1--2.60~$\upmu$m. These data show fluctuations of order tens of degrees across wavelength scales of order 0.1~$\upmu$m, with a mean angle of $160\pm4^\circ$. These sub-structures may reflect variations in the photospheric pressures probed by the data, as deeper pressures will tend to probe hotter temperatures, whose emission will trend shorter in wavelength. This points to the potential for hot spots to explain the variability on the largest wavelength scales. Valid fits for sinusoid models end at approximately 2.60~$\upmu$m, with two isolated exceptions, and resumes at 3.36~$\upmu$m within the previously described 3.19--3.44~$\upmu$m feature. We divide the remaining data into two ranges: 3.44--4.35~$\upmu$m and 4.35--5.30~$\upmu$m. 4.35~$\upmu$m as a dividing line was chosen based on where the best-fit variability amplitudes begin to exceed 2\%. These two clusters are distinguished by their mean variability amplitudes, at $1.64\pm0.05$\%\ and $2.47\pm0.07$\%, and are also distinct in their mean phases, at $143\pm3^\circ$ and $82\pm2^\circ$, respectively. One interpretation, if the observed variability is indeed linked to rotation, is that the origin of the variability at wavelengths greater than 4.35~$\upmu$m is spatially distinct from that of the variability at 1--2.60~$\upmu$m, and possibly that for the 3.44--4.35~$\upmu$m data.

\begin{figure}
\begin{center}
\includegraphics[width=8.5cm]{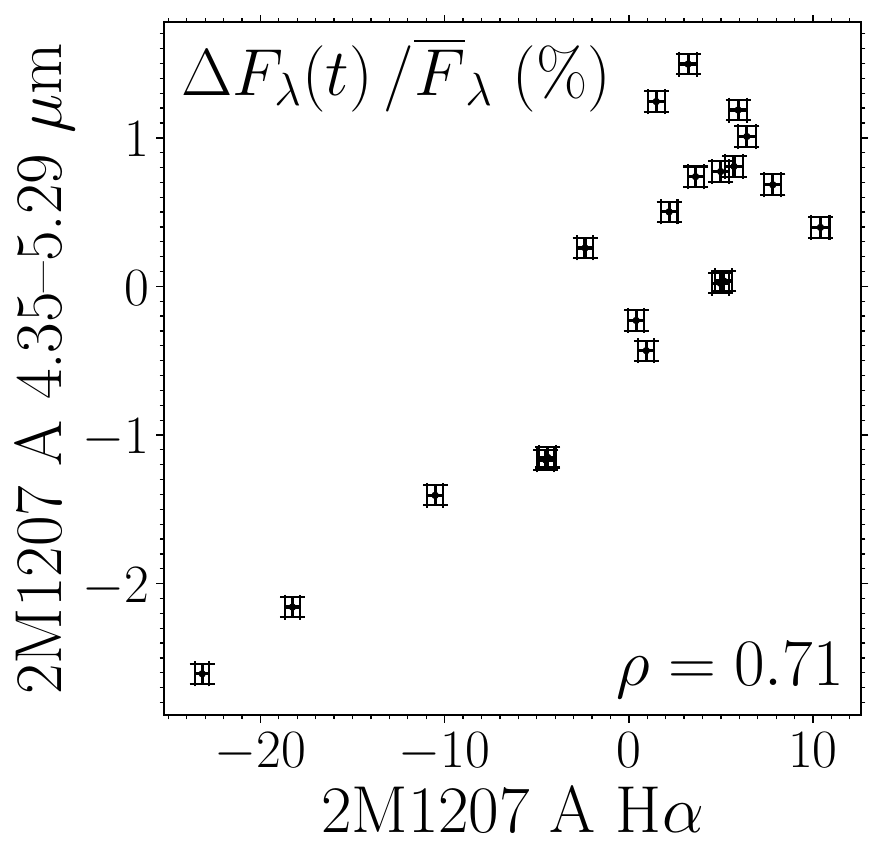} \\
\includegraphics[width=8.5cm]{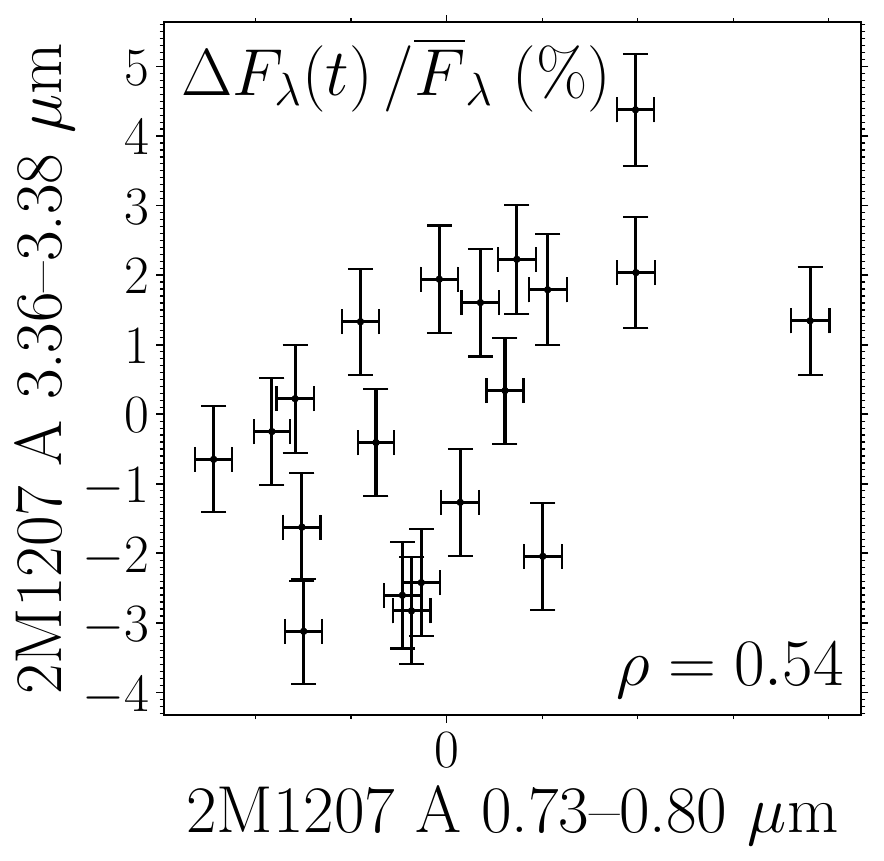}
\caption{A comparison of the variability displacements across time of the two pairs of light curves shown in Figure \ref{fig:variability-map-2M1207A}. \emph{Top:} The variability amplitudes across time for the H$\alpha$-sensitive wavelength elements (0.65--0.66~$\upmu$m) with those $>4.35$~$\upmu$m. We can evaluate the co-variance of the light curves in time with a Spearman rank correlation coefficient, which returns a value of approximately 0.71. \emph{Bottom:} The variability amplitudes across time for light curves at 0.73--0.80 and 3.36--3.38 $\upmu$m, two regions whose sinusoidal trend models return mutually consistent parameter values.}
\label{fig:2M1207A_variability_displacement_comparisons}
\end{center}
\end{figure}

\subsubsection{Interpreting the Variability in Localized Wavelength Regions}\label{sec:discussion:A:light-curves}
We now discuss the potential interpretations of the two light curve pairs presented in Figure~\ref{fig:variability-map-2M1207A}d and e. Starting with (d), the retrieved phases in the three wavelength elements that are sensitive to H$\alpha$ emission (0.65--0.66~$\upmu$m) are distinct from the fits to neighboring wavelength elements, more closely matching the average phase of the 4.35--5.30~$\upmu$m data. The mean phase of the H$\alpha$ fits is $82\pm20^\circ$, consistent with the phase derived from the data beyond 4.35~$\upmu$m. One interpretation is that the sources of variability in both the H$\alpha$ emission and the $>4.35$~$\upmu$m data are temporally (and therefore potentially spatially) coupled. While the amplitudes of the two ranges differ considerably, their light curves appear to be correlated in time (see the left panel of Figure~\ref{fig:2M1207A_variability_displacement_comparisons}). We evaluate the co-variance of the light curves in time with a Spearman rank correlation coefficient; this returns a value $\rho=0.71$, indicating a moderate to strong correlation.

We consider potential physical interpretations of these features. 2M1207~A's effective temperature of approximately 2600 K places it as a late M spectral type, which means that variability is unlikely to be primarily driven by the time evolution of clouds. The distinct phase of the reddest data --- those beyond 4.35~$\upmu$m --- suggest there may be some material associated with 2M1207~A at a temperature considerably colder than the photosphere. As mentioned previously, analyses with Spitzer time series suggest disk asymmetries can contribute to spectral variability; therefore, the disk around 2M1207~A provides a plausible source for such material. Additionally, since H$\alpha$ is known to be a tracer of accretion\footnote{We note that H$\alpha$ emission is not exclusive to accretion, which introduces a minor caveat; see e.g.~\citet{Viswanath2024}.}, this colder material may be linked in time/space with a region on the primary that is in an active episode of accretion, that then rotates out of view across the observation. Unfortunately, given the low spectral resolution of our data (R $\approx 54.5$ at 0.656~$\upmu$m), we are unable to get a precise constraint on the velocity broadening in the H$\alpha$ feature, limiting further analysis.

The second pair of light curves, shown in Figure~\ref{fig:variability-map-2M1207A}e, are integrated across 0.73--0.80 and 3.36--3.38 $\upmu$m, respectively. The best-fit sinusoidal model parameters for these light curves are consistent with each other. Their apparent time correlation however is a bit weaker, at $\rho=0.54$, and show a relative offset in time of order one integration time. The source of the variability at 3.36--3.38 $\upmu$m is not immediately clear. This feature lies within the broad 3.19--3.44~$\upmu$m range that shows more rapid variability than surrounding wavelengths; many of the wavelength elements in this range lack well-constrained sinusoidal fits. We are therefore unable to comment on the potential spatial distribution of this variability feature. In cooler objects (2M1207~b notwithstanding) a known vibrational mode of CH$_4$ is often seen at $\approx 3.3$~$\upmu$m. The width of our feature in wavelength space is similar to that of the trough caused by the $P$, $Q$, and $R$ branches of CH$_4$ --- a trough that can be seen in, for example, the spectrum of the T dwarf GJ 229 B \citep[e.g.][]{Burrows1999,Howe2022}. However, 2M1207~A's effective temperature of 2600 K does not support the presence of CH$_4$; a range $T_\mathrm{eff}\approx1200$--1500~K is consistent with previous theoretical predictions of the onset of CH$_4$ in sub-stellar spectra \citep[see e.g.][]{Burrows1999,Kirkpatrick1999,Kirkpatrick2005,Beiler2022}.\footnote{Recently, a new spectral type H has been proposed in \citet{Luhman2025} which includes young sub-stellar objects that show absorption features centered at 3.4~$\upmu$m that are associated with hydrocarbons, most often attributed to CH$_4$. These are seen despite their spectral shape and features otherwise being classified as early as the M/L transition.} Recent MIRI data of 2M1207~A, as reported in \citet{Patapis2025}, show that the circum-stellar disk is rich in hydrocarbons; therefore, it is possible for this source of variability to be related to this disk material. The correlation with the 0.73--0.80 $\upmu$m may be due to material in the upper atmosphere interacting with the disk. However, we do not have strong enough evidence to comment further.

Finally, we discuss narrow features in the variability map which occur at approximately 1.9 and 3.9 $\upmu$m; these features may not have an origin in 2M1207~A itself. These are likely instead to be artifacts in the flux subtraction. When reviewing the equivalent variability maps made with a previous CRDS context, these features do not persist. No clear outliers in the time-resolved data can be identified, such as a spurious pixel value --- therefore, we exclude these features from our analysis.

\subsection{Interpreting the Variability in 2M1207~b}
\label{sec:discussion:b}
We begin with a cursory interpretation of the variability in H$\alpha$ emission, as well as in the 0.86--1.77 and 3.29--4.34~$\upmu$m regions. Since we only find support for linear trends, this suggests that, if these trends are due to rotation, that we do not fully capture the rotation period within the observing window. Additionally, the variability at 3.29--4.34~$\upmu$m does not show a gradual increase; instead, most of the increase in flux occurs between the two exposures that occur between hours 5 and 6 of the observation. The flux in the 0.86--1.77~$\upmu$m region crosses its mean value within the same interval. One physical interpretation of the shape of these two light curves is that the feature that generates the variability in the redder region may be more concentrated spatially than that of the bluer region, yielding a sharper change in time. However, we are limited in how strongly we can argue for specific physical mechanisms; our caveats are explained below.

\subsubsection[Evaluating the Potential for Contamination in Halpha]{Evaluating the Potential for Contamination in H$\alpha$}\label{sec:discussion:b:Halpha_contamination}
\begin{figure}
\begin{center}
\includegraphics[width=8.5cm]{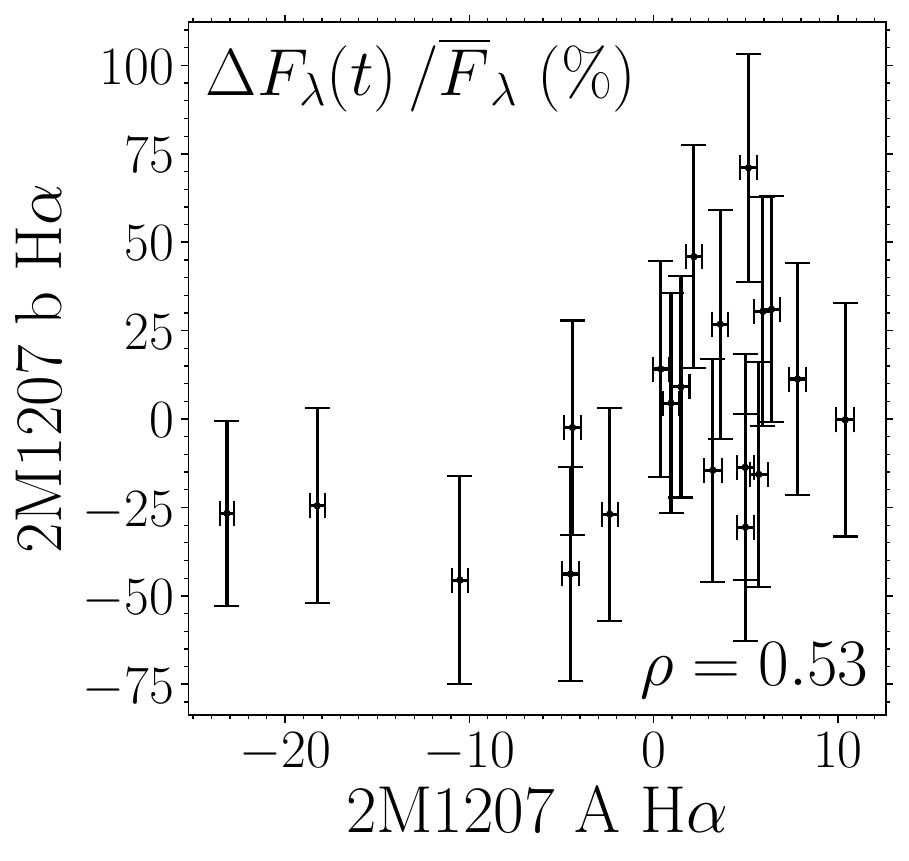}
\caption{A comparison the variability displacements of the H$\alpha$ light curves between 2M1207~A and b. The correlation coefficient between the two light curves, which probes the potential for contamination in the signals between the objects, is approximately 0.53.}
\label{fig:Halpha_variability_displacement_comparison}
\end{center}
\end{figure}

As noted in \S \ref{sec:observations:post_subtraction_residuals} and \ref{sec:variability-models:b:error-estimates},  the sampled background fluxes, which are taken as a pessimistic estimate of our spectral uncertainties for 2M1207~b, sampled show a flux level close to the companion's up through $\sim 1$~$\upmu$m. One feature in this wavelength is H$\alpha$ feature which is detected albeit only at a SNR~$\sim 7$. To examine the potential for contamination in the H$\alpha$ feature from that of 2M1207~A, we employ a Spearman rank correlation test, which compares the co-variation of 2M1207~b's H$\alpha$ light curve in time with that of 2M1207~A. We find a correlation coefficient of $\rho=0.53$ (Figure \ref{fig:Halpha_variability_displacement_comparison}), which indicates a moderate correlation in time; therefore, while there may be some contamination, it does not explain all of the observed signal.

\subsubsection{Broad Agreement with Previous Results}
\label{sec:discussion:b:model_comparison}

In \S \ref{sec:atmosphere-models} we showed that the absolute scale of the cloud filling fractions is lower across all times, at 7.12--7.59\%, versus the 8.8\% originally constrained in \citet{Zhang2025}. Our interpretation of this difference is limited due to the lack of an independent full-parameter retrieval; however, this invites the possibility that 2M1207~b may show long-term variability in its cloud covering fractions. Despite the absolute difference, our results show that the cloud variability mechanism in the atmospheric model is broadly consistent with the variability amplitudes measured from our data (see Figure \ref{fig:ZJ_forward-model-spectra_variability}). To examine this further, we compare these amplitudes with a focus on the two wavelength regions (0.86--1.77 and 3.29--4.34~$\upmu$m) showing the most significant variability. Note that these are not the statistics of the absolute atmospheric model fluxes versus the data, but rather a comparison of their predicted variability amplitudes. Fitting the variability model to data-derived variability amplitudes in the 0.86--1.77~$\upmu$m region yields $\chi_r^2=2.2$. The fit to the 3.29--4.34~$\upmu$m region results in a $\chi_r^2=0.9$, indicating a good fit where the data do not necessitate further model complexity. Overall, this preliminary comparison suggests that the interpretation of \citet{Zhang2025} regarding L dwarf variability driven by two distinct cloud components holds broadly, with some isolated exceptions. This agreement is notable given that our overall model spectrum fit is not yet fully optimized to our data. Building upon this initial assessment, and to account for the fact that our fits to the atmospheric parameters of each column are not yet optimized, we use the 0.86--1.77~$\upmu$m fit as a baseline to evaluate potential ``excess'' disagreement at other wavelengths. To normalize the $\chi_r^2$ of our baseline region to unity, we inflate the error bars by $\sqrt{2.2}\approx1.5$. We identify and discuss possible interpretations of two isolated areas of disagreement using this metric.

At wavelengths $<1.1$~$\upmu$m, we find $\chi_r^2=4.4$ without error inflation, which corresponds to a $\approx 2.1 \sigma$ excess inconsistency with the model. Note that the \citet{Zhang2025} retrievals rely on data with a lower wavelength limit of 0.97~$\upmu$m, so the comparison at shorter wavelengths is novel. Our data begin to diverge from the model predictions very close to this wavelength cutoff, suggesting either our data are too noisy below this wavelength, or that an additional model component is needed to explain the 0.86--0.97~$\upmu$m variability amplitudes. Both iron and silicate clouds strongly suppress the outgoing flux at wavelengths $\lesssim 1$~$\upmu$m --- therefore, the source of this additional variability amplitude is unlikely to originate from a variation in the relative covering fraction of these condensates. Our analysis in \S~\ref{sec:discussion:b:Halpha_contamination} raises the possibility that these variability amplitudes are confounded with a signal from the primary. If the variability amplitudes $\lesssim 1$~$\upmu$m represent a true signal from 2M1207~b, one possibility is that we are seeing the effects of ``rainout'' chemistry, which affects molecules with significant absorption $\lesssim 1$~$\upmu$m such as FeH, TiO, and VO \citep[e.g.][]{Fegley1994,Burrows2001,Lodders2002,Sharp2007}. Rainout can alter the observed abundances of these species relative to predictions from chemical equilibrium. \citet{Rowland2023} have recently demonstrated through model retrievals that, due to rainout, FeH in early- to mid-L dwarfs can drop in abundance by two orders of magnitude at pressures shallower than $\sim 1$~bar. They also remark that TiO may have strongly non-uniform abundances at temperatures as cool as 1400~K. Therefore, temperature fluctuations may induce variability through changes in the abundance of molecules such as TiO at 0.86--0.97~$\upmu$m. Additional data at shorter wavelengths would be helpful to investigate this hypothesis.

The second area of disagreement is 1.9--2.4~$\upmu$m. The potential source of disagreement is less clear. However, the amount of disagreement is also less severe, with $\chi_r^2=1.9$. If the previous error inflation were applied, these data would agree with the model at least as well as the 0.86--1.77~$\upmu$m data overall (i.e. $\chi_r^2 \lesssim 1$). In the current atmospheric models, the current lever arm controlling the variability at these wavelengths is the observed column depth of H$_2$O, which is modulated by the changes in the cloud covering fraction. It is unlikely the data we present here will be able to significantly refine the compositional constraints made on cloud species, given the relatively low spectral resolution when compared with the previously characterized JWST spectrum. Such a refinement is more likely to come from data in the mid-infrared, where condensates such as silicates show a strong signal. However, the extension of data below 0.97~$\upmu$m suggests that follow-up atmospheric characterization studies may consider testing to what extent fluctuations in abundances of rainout chemicals could influence the spectrum at these wavelengths.

%%%%%%%%%%%%%%%%%%%%%%%%%%%%%%%%%%%%%%%%%%%%%%%
\section{Conclusions}
\label{sec:conclusions}
%%%%%%%%%%%%%%%%%%%%%%%%%%%%%%%%%%%%%%%%%%%%%%%
This work presents new JWST NIRSpec/PRISM IFU observations of the 2M1207 system, comprised of 20 time-resolved spectra for both objects across a 12.56-hour span. We find:
\begin{enumerate}
    \item We measure a $\mathrm{SNR} \sim 100$ per wavelength bin per time for 2M1207~A's spectra and variability semi-amplitudes of order $\pm 2$\%. H$\alpha$ emission is detected at a $\mathrm{SNR} \approx 10$ with a variability amplitude that spans a range of $\approx 30$\%\ across the observation.
    
    \item A large fraction of the spectrum of 2M1207~A shows statistical evidence for non-linear trends in its time series, which we model as a signal from its rotation via a sinusoid. However, we only place a lower limit on the rotation period, as the inferred period constraints almost all exceed the observing duration.
    
    \item The retrieved phase angles of the above sinusoidal models for 2M1207~A suggest that the maximum fluxes tend to occur later with decreasing wavelength. The variability signal beyond 4.35~$\upmu$m is moderately temporally correlated with that of H$\alpha$, as well as a pair of light curves at 0.73--0.80 and 3.36--3.38~$\upmu$m. This suggests that accretion-sensitive variability may be linked spatially to material --- either within the circum-stellar disk or in the atmosphere --- significantly cooler than 2M1207~A's photospheric temperature of 2600~K.
    
    \item We measure a SNR $\sim 185$ across most of each of 2M1207~b's time-resolved spectra, with variability semi-amplitudes of order $\pm 0.5$\%, increasing to $\pm 2$\%\ near the $J$ band peak. The data below approximately 0.86~$\upmu$m have a $\mathrm{SNR} \lesssim 1$. We determine that 2M1207~A may contaminate the signal of 2M1207~b at the bluest wavelengths, with noise levels comparable to 2M1207~b's flux $\lesssim 1$~$\upmu$m, then decreasing below 10\% of the flux by $\approx 1.3$~$\upmu$m.
    
    \item An empirical characterization of 2M1207~b's spectral variability reveals two wavelength ranges, at 0.86--1.77~$\upmu$m and 3.29--4.34~$\upmu$m, that do not provide evidence for periodic variations, but do support a linear model which may represent either a non-periodic signal or a first-order approximation to a sinusoid. This limits our interpretation of the variability data as a robust detection of rotational modulation, but does not rule it out entirely.
    
    \item By re-fitting an atmospheric model used to characterize previous 2M1207~b JWST data, we demonstrate that these wavelength regions are broadly consistent with the previous interpretation of two types of surface areas with different cloud properties, whose relative area varies with time.
    
    \item One exception to the above is the data below 0.97~$\upmu$m, which were not available in previous JWST data and diverge from the model predictions by $\approx 4 \sigma$. As mentioned previously, this area of the spectrum may be contaminated by the signal from 2M1207~A. If a signal from 2M1207~b itself, one potential source is rainout chemistry, which can significantly alter the shape of the spectrum at $\lesssim 1$~$\upmu$m.m.
\end{enumerate}
This work demonstrates that JWST is capable at low brightness contrasts to simultaneously monitor the variability of planetary-mass companions and their hosts spectroscopically. The spectral variability of 2M1207~b broadly supports the picture of a young exoplanet with a heterogeneous atmosphere, displaying spatially distinct cloud structures consistent with the banded structures inferred from long-term variability monitoring campaigns. The robust detection of H$\alpha$ variability in 2M1207~A, and tentative evidence for its time correlation with near-infrared variability features, provide a step toward a comprehensive time-resolved picture of brown dwarfs and their circumstellar environments at early stages. These data will benefit from a more comprehensive atmospheric characterization of both 2M1207~A and b. The data would also be well complemented by additional variability monitoring over multiple rotations --- as well as time-resolved data in the mid-infrared --- to reveal its weather and climate more completely.

\bibliography{library}{}
\bibliographystyle{aasjournalv7}

\appendix

%%%%%%%%%%%%%%%%%%%%%%%%%%%%%%%%%%%%%%%%%%%%%%%
\section{A New Astrometric Measurement Provides a Weak Constraint on Orbital Inclination}
\label{sec:appendix:astrometry}
\begin{figure*}
\begin{center}
\includegraphics[width=17cm]{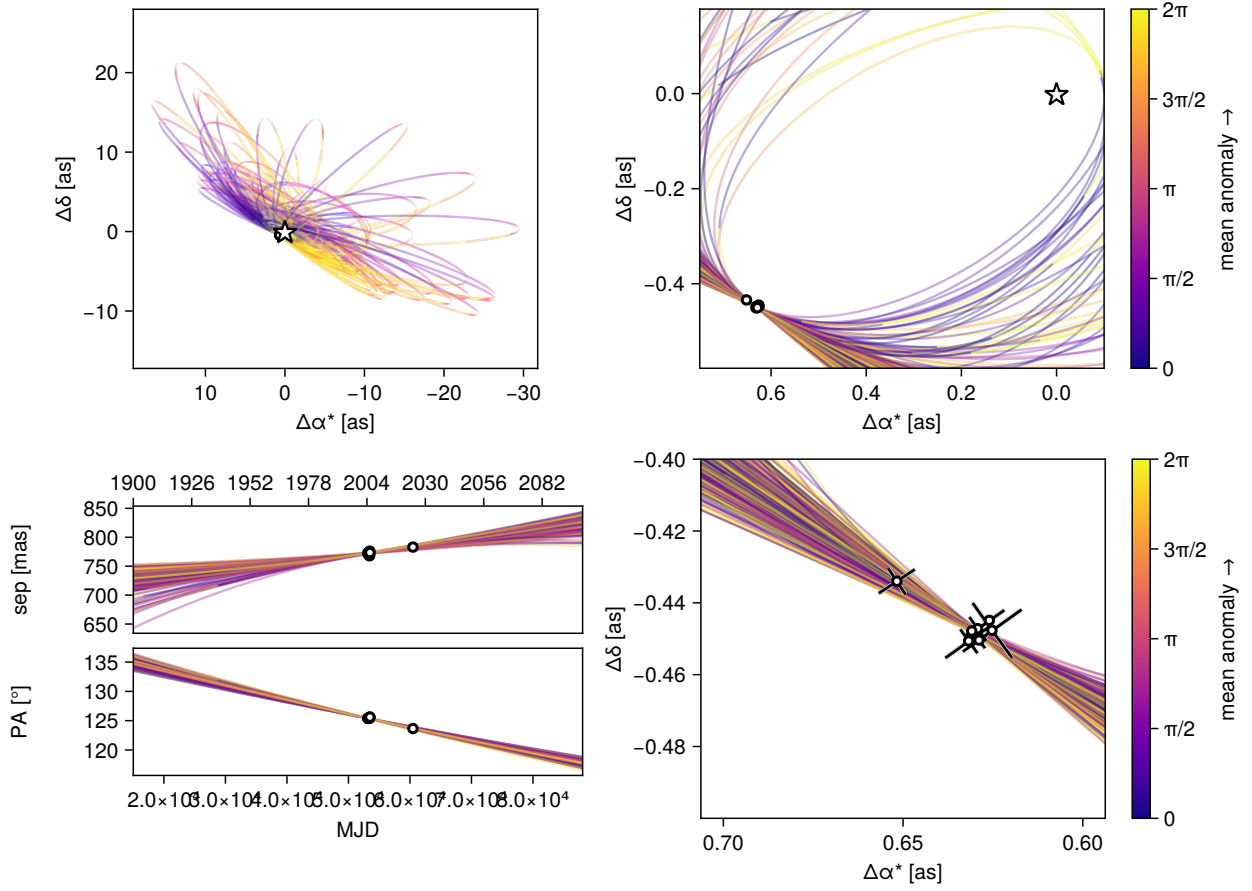}
\caption{The distribution of orbital solutions for the position angles and angular separation data for 2M1207~b. The color denotes the mean anomaly across each orbit. The plots in the right column zoom in more closely to the data; the isolated position parameters are shown in the bottom-left panel.}
\label{fig:astrometric-fit_orbit-plots}
\end{center}
\end{figure*}

%%%%%%%%%%%%%%%%%%%%%%%%%%%%%%%%%%%%%%%%%%%%%%%
We use \texttt{octofitter} \citep{Thompson2023} to fit seven pairs of angular separations and position angles for the companion: six data points observed from 2004--2005, taken from the literature \citep{Chauvin2005a,Song2006,Mohanty2007}, as well as a new constraint from our observations. The new measurement is compiled with previous measurements in Table \ref{table:companion_astrometry}.\footnote{An additional astrometric measurement is possible using HST observations from 2011 \citep[Program ID: 12225, PI: A.\ Reiners, see also][]{Marleau2024}. However, the uncertainties in the position angles and separations are expected to be a factor of several higher compared with the data presented here.} The fits shown in Figure~\ref{fig:astrometric-fit_orbit-plots} use \texttt{octofitter}'s built-in MCMC fitting routine, \texttt{octofit}, which employs a Hamiltonian Monte Carlo algorithm with No U-Turns Sampling \citep[referred to as \texttt{AdvancedHMC} with NUTS in the code; see][]{Hoffman2014,Xu2020}. We keep the default settings for this version of the sampler with the exception of a target proposal acceptance rate of 95\%, and running for $10^5$ iterations with an initial adaptation stage of $10^4$ iterations. The run is initialized by selecting the highest posterior density sample out of 250,000 samples randomly generated from the priors which are summarized in Table \ref{table:companion_astrometry_fits}. In short, we impose broad priors with the exception of the parallax, which uses the GAIA-derived distance estimate of $64.5\pm0.5$ pc \citep{Bailer-Jones2021,Luhman2023}. In addition to these priors, we also tested fits to the data with a wide prior in parallax (uniform across 1--100 mas), and with a tight prior on the total system mass (a normal prior with $\mu=25.1$, $\sigma=6.3$ $M_\mathrm{Jup}$, truncated at a lower bound of 1 $M_\mathrm{Jup}$).

\begin{figure*}
\begin{center}
\begin{tabular}{cc}
\includegraphics[width=17cm]{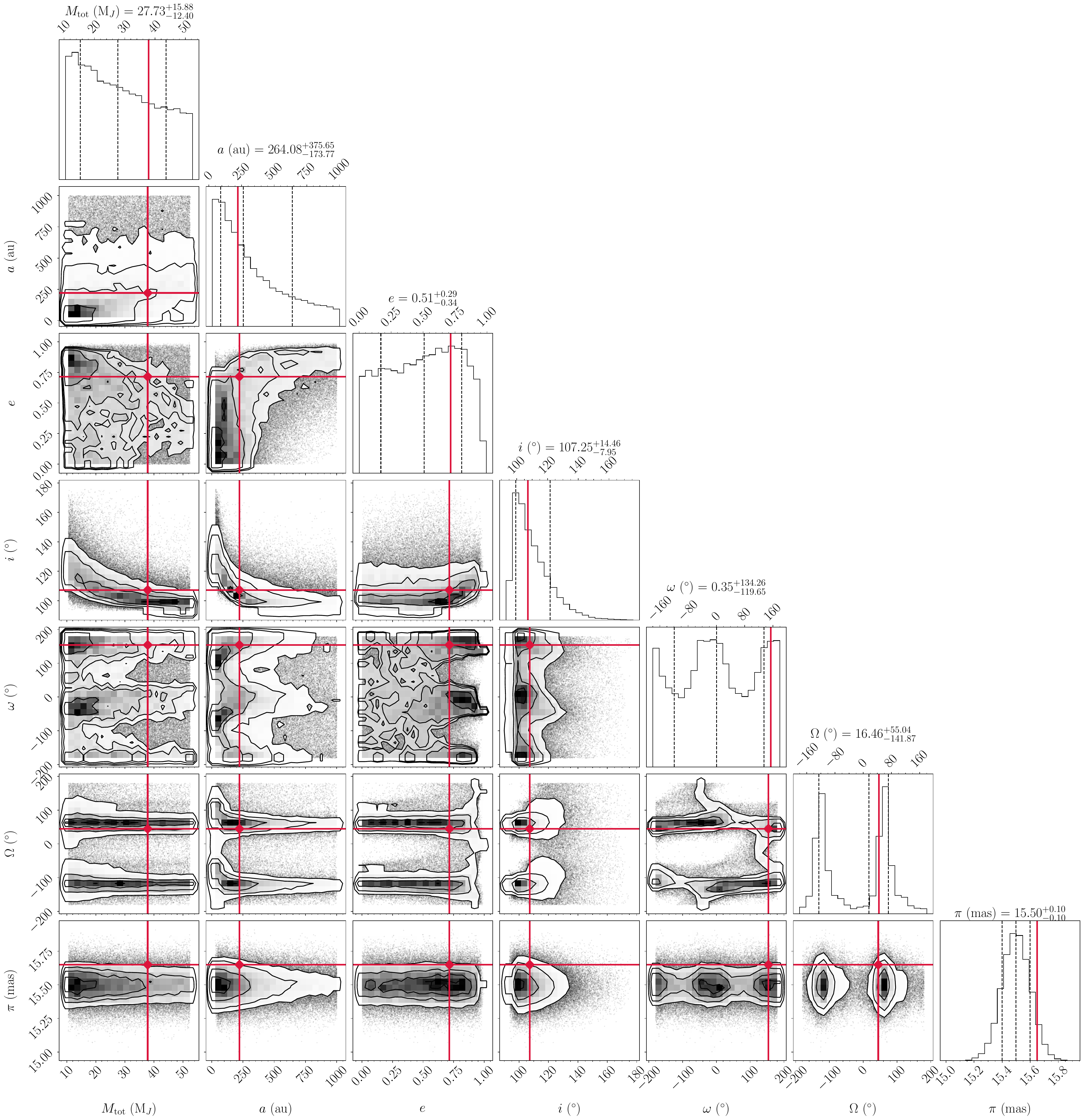} &
\hspace{-6.5cm}\raisebox{10cm}{\includegraphics[width=6.5cm]{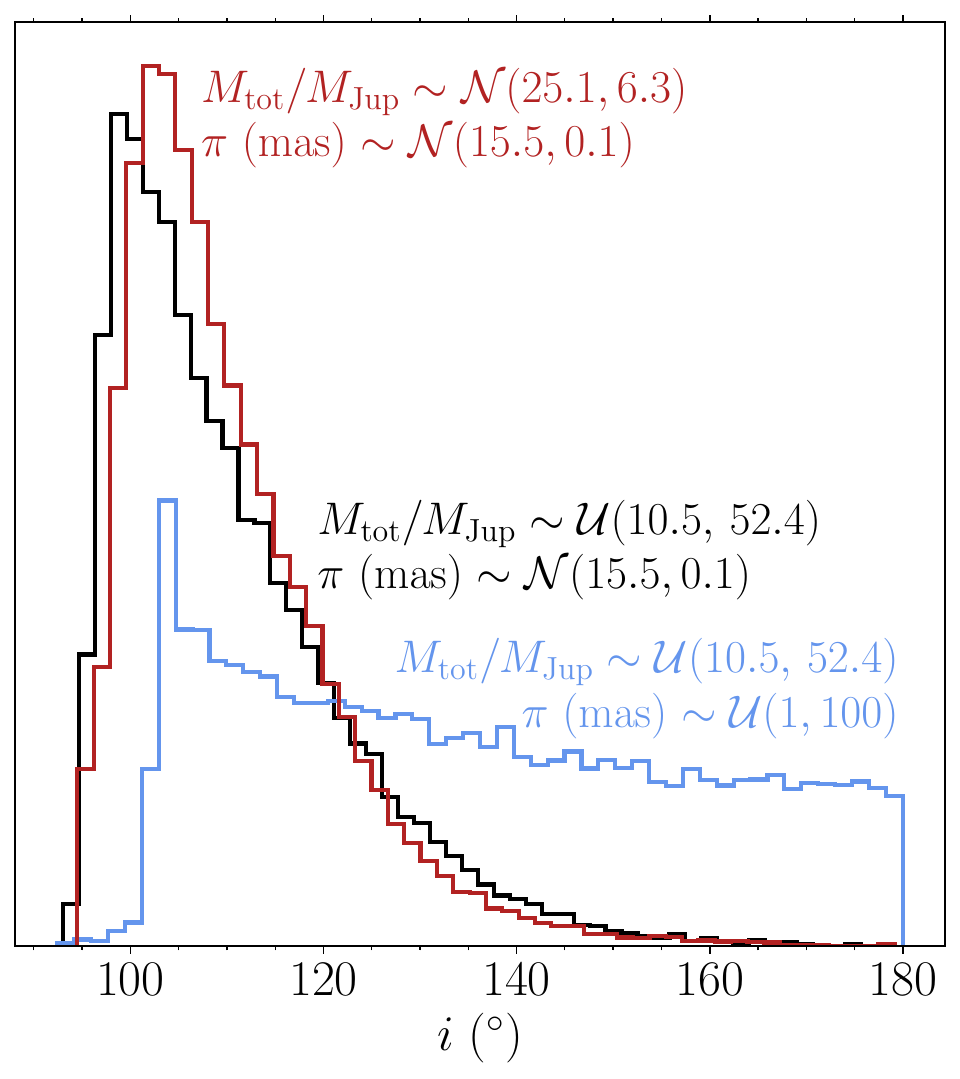}}
\end{tabular}
\caption{The retrieved posterior distributions for the \texttt{Octofitter} fit to the available position angles and angular separations available in the literature for 2M1207, including a new datum presented in this work. The maximum-likelihood estimates of the orbital parameters is shown with the red point and cross-hairs in each sub-plot. In the top-right plot we show a comparison of the posterior distributions for the orbital inclination for three choices of priors. The black histogram shows the ``base'' run with the priors shown in Table \ref{table:companion_astrometry_fits}. The blue histogram shows a run with all priors the same except for a broader prior in parallax (1--100 mas). The red histogram shows a run with a prior on the total system mass taken from a combination of constraints from \citet{Chauvin2005a} and \cite{Zhang2025}.}
\label{fig:astrometric-fit_corner-plot}
\end{center}
\end{figure*}

Figures \ref{fig:astrometric-fit_orbit-plots} and \ref{fig:astrometric-fit_corner-plot} show that the overall shape and orientation of the orbit are not precisely constrained, as indicated by the broad ranges of the posterior distributions of our orbital parameters. The strongest constraint is that on the orbital inclination, driven by the new datum which constrains the orbital motion to inclinations between 90--$180^\circ$, with a 68\% credible interval of $i={107^\circ}^{+14}_{-8}$. When comparing the posterior distributions from the two other runs (top-right panel of Figure \ref{fig:astrometric-fit_corner-plot}), we see that the GAIA-derived parallax prior is responsible for the strong peak at inclinations close to edge-on. The 68\% credible interval for the 1--100 mas parallax prior is $133^\circ{}^{+31}_{-23}$. The tighter prior on mass does not provide much additional change to the constraint, with a credible interval $108^\circ{}^{+12}_{-7}$.

We compare these results with existing constraints of the orientation of both 2M1207~A's disk and of its outflow. While the position angle of the disk is constrained to $174^\circ\pm12$ in \citet{Ricci2017}, there remains a large uncertainty in the disk inclination, with constraints of $i=70$--$75^\circ$, 57--$69^\circ$, and 20--$55^\circ$ from \citet{Skemer2011}, \citet{Riaz2012a}, and \citet{Ricci2017}, respectively. This limits us in making a more accurate constraint on the true relative orientation of the orbital plane with the disk. The \citet{Ricci2017} result also yielded a dynamical mass constraint $M_\mathrm{A} = 60^{+80}_{-20} M_\mathrm{Jup}$, substantially higher than the masses constrained from evolutionary models. In their disk model, lower masses would imply a disk inclination closer to edge-on. This lack of a precise disk inclination limits a more accurate constraint on their relative orientation. Relative to the outflow, if we assume the vector is very nearly in the plane of the sky \citep[][]{Whelan2012}, the outflow is consistent with being aligned with the orbital plane, with an offset of $0^\circ\pm22$. However, note that the uncertainties are large, and that this range is a lower limit of the uncertainties given we assume the outflow is completely aligned with the plane of the sky. This limits the extent to which we can physically interpret these results.

\begin{deluxetable}{rccc}[htb!]\label{table:companion_astrometry}
\tabletypesize{\normalsize}
\tablewidth{0pt}
\tablecaption{Compilation of astrometric measurements of 2M1207~b from the literature. ``C05'' refers to \citet{Chauvin2005a}, ``S06'' refers to \citet{Song2006}, and ``M07'' refers to \citet{Mohanty2007}.}
\tablehead{
    \colhead{Epoch} &
    \colhead{Separation (mas)} &
    \colhead{Position Angle (${}^\circ$)} &
    \colhead{Reference}
    }
    
\startdata
2004.32 & $772\pm4$ & $125.4\pm0.3$ & C05 \\
2004.66 & $773.7\pm2.2$ & $125.37\pm0.03$ & S06 \\
2005.10 & $768\pm5$ & $125.4\pm0.3$ & C05 \\
2005.23 & $769\pm10$ & $125.6\pm0.7$ & M07 \\
2005.24 & $776\pm8$ & $125.5\pm0.3$ & C05 \\
2005.32 & $773.5\pm2.3$ & $125.61\pm0.20$ & S06 \\
2024.49 & $773.7\pm2.2$ & $123.66\pm0.32$ & This work \\
\enddata
\end{deluxetable}

\begin{deluxetable}{rccc}[htb!]\label{table:companion_astrometry_fits}
\tabletypesize{\normalsize}
\tablewidth{0pt}
\tablecaption{The results of our astrometric fit. $\mathcal{U}$ denotes a uniform prior with lower and upper bounds, respectively, while $\mathcal{N}$ denotes a normal prior with mean and scale. The Sine prior is uniform in the sine of the angle across the given range. The \texttt{UniformCircular} prior calculates normal priors on two Cartesian coordinates in the plane in which the angle is defined, then converts to an angle prior that smoothly wraps around the domain 0--$2\pi$ radians. Alternate runs tested a wide parallax prior ($\mathcal{U}\!\left(1,\,100\right)$ mas) and a mass prior derived from evolutionary models \citep[$\mathcal{N}\!\left(25.1,\,6.3\right)$ $M_\mathrm{Jup}$; see][]{Chauvin2005a,Zhang2025}.}
\tablehead{
    \colhead{Parameter} &
    \colhead{Prior} &
    \colhead{MLE Value} &
    \colhead{Median and 68\% Credible Interval}
    }
    
\startdata
$M_\mathrm{A}+M_\mathrm{b}$ ($M_\mathrm{J}$)    & $\mathcal{U}\!\left(10.5,\,52.4\right)$   & 38    & $28^{+16}_{-12}$ \\
$a$ (au)                                        & $\mathcal{U}\!\left(0,\,1000\right)$      & 221   & $264^{+376}_{-174}$ \\
$e$                                             & $\mathcal{U}\!\left(0,\,1-10^{-4}\right)$ & 0.72  & $0.51^{+0.29}_{-0.34}$ \\
$i$ ($^\circ$)                                  & Sine$\left(0,\,180\right)$                & 107   & $107^{+14}_{-8}$ \\
$\omega$ ($^\circ$) & \texttt{UniformCircular}$\left(0, 360\right)$ & 154 & $0^{+134}_{-120}$ \\
$\Omega$ ($^\circ$) & \texttt{UniformCircular}$\left(0, 360\right)$ & 45 & $16^{+55}_{-142}$ \\
$\pi$ (mas) & $\mathcal{N}\!\left(15.5,\,0.1\right)$ & 15.6 & $15.5\pm0.1$\tablenotemark{a} \\
\enddata
\tablenotetext{a}{The posterior here is almost entirely shaped by the prior.}
\end{deluxetable}

%%%%%%%%%%%%%%%%%%%%%%%%%%%%%%%%%%%%%%%%%%%%%%%
\section{Atmospheric Model Configuration for 2M1207~b}
\label{sec:appendix:2M1207b_atmospheric_model_configuration}
%%%%%%%%%%%%%%%%%%%%%%%%%%%%%%%%%%%%%%%%%%%%%%%
Both the forward models and retrievals in the \citet{Zhang2025} study are performed with \texttt{petitRADTRANS} \citep[abbreviated \texttt{pRT}, see][]{Molliere2020,Nasedkin2024a}.
The \texttt{QEQ-1} model makes the following assumptions:
\begin{itemize}
    \item The atmospheric composition is assumed to be in chemical equilibrium, with two exceptions described below. Equilibrium abundances are calculated at a given temperature-pressure (T--P) profile, surface gravity, carbon-to-oxygen (C/O ratio), and metallicity. The code uses the \texttt{easychem} chemical equilibrium model \citep{Lei2024} to perform these calculations.
    \item The first exception to chemical equilibrium is: at pressures shallower than a ``quenching'' pressure, the principal molecules involved in the carbon chemistry --- CO, CH$_4$, and H$_2$O --- retain their mass mixing ratios as set by the equilibrium values at that quenching pressure. The quenching pressure is treated as a free parameter.
    \item There exist two cloud species (forsterite silicates, i.e. Mg$_2$SiO$_4$, and iron, Fe), each condensing into a single contiguous cloud layer. Both species are modeled with Mie scattering; the forsterite opacities assume amorphous grains, while the iron opacities assume crystalline grains.
    \item The second exception to chemical equilibrium is: the mass mixing ratio of each cloud species is set by a parametrization originally described in \citet{Ackerman2001}. The form implemented here is slightly modified from the original function, as
    \begin{equation}\label{eq:ackerman-marley-clouds}
    X\!\left(P\right) = X_\mathrm{base} \left(\frac{P}{P_\mathrm{base}}\right)^{f_\mathrm{sed}}
    \end{equation}
    where $X$ denotes the mass mixing fraction of the condensate. The base pressure $P_\mathrm{base}$ is set by the intersection of the condensation curve with the T--P profile.
    \item The emission spectrum assumes two distinct components, representing spatially distinct atmospheric surface areas (hereafter ``columns''). These columns are configured to share the same T--P profile, chemistry, and cloud species; but the cloud properties within each column are allowed to vary independently. The composite spectrum is taken to be a normalized linear combination of the spectra generated from these two columns.
\end{itemize}
A few additional parameters related to the cloud models are needed to fully reproduce the emission spectra. The full configuration is summarized in Table \ref{table:pRT-parameters}.

\begin{deluxetable}{rc}[htb]\label{table:pRT-parameters}
\tabletypesize{\normalsize}
\tablewidth{0pt}
\tablecaption{Parameter values for the statistically-preferred atmospheric model as reported in \citet{Zhang2025}. The reported values for $\log_{10} K_\mathrm{zz}$ and $\sigma_\mathrm{g}$ are taken from private communication with the lead author. Both cloud species are modeled with Mie scattering; the forsterite opacities assume amorphous grains, while the iron opacities assume crystalline grains.}
\tablehead{
    \colhead{Parameter} &
    \colhead{Fit Value}
    }
\startdata
$R / R_\mathrm{Jup}$                        & 1.399     \\
$\log g$ (dex, cm s$^{-2}$)                   & 3.62      \\
$\left[\mathrm{M}/\mathrm{H}\right]$ (dex)  & $-0.05$   \\
C/O Ratio                                   & 0.440     \\
$\log_{10} P_\mathrm{quench}$ (bar)         & $-3.3$    \\
\cutinhead{\textbf{Cloud column 1: ``Thin clouds''}}
Covering fraction (\%)                      & 8.65      \\
$\log_{10} K_\mathrm{zz}$ (cm$^2$ s$^{-1}$) & 5.376     \\
$\sigma_\mathrm{g}$                         & 1.473     \\
\multicolumn{2}{l}{\textit{Forsterite (Mg$_2$SiO$_4$)}} \\
\hline \\
$\log_{10} X_\mathrm{base}$ (dex)           & $-7.77$   \\
$f_\mathrm{sed}$                            & 4.54      \\
\multicolumn{2}{l}{\textit{Iron (Fe)}} \\
\hline \\
$\log_{10} X_\mathrm{base}$ (dex)           & $-4.53$   \\
$f_\mathrm{sed}$                            & 4.53      \\
\cutinhead{\textbf{Cloud column 2: ``Thick clouds''}}
Covering fraction (\%)                      & 91.35     \\
$\log_{10} K_\mathrm{zz}$ (cm$^2$ s$^{-1}$) & 8.962     \\
$\sigma_\mathrm{g}$                         & 1.146     \\
\multicolumn{2}{l}{\textit{Forsterite (Mg$_2$SiO$_4$)}} \\
\hline \\
$\log_{10} X_\mathrm{base}$ (dex)           & $-0.67$   \\
$f_\mathrm{sed}$                            & 1.20      \\
\multicolumn{2}{l}{\textit{Iron (Fe)}} \\
\hline \\
$\log_{10} X_\mathrm{base}$ (dex)           & $-4.84$   \\
$f_\mathrm{sed}$                            & 5.59      \\
\enddata
\end{deluxetable}

\begin{acknowledgements}
This work is based on observations made with the NASA/ESA/CSA James Webb Space Telescope. The data were obtained from the Mikulski Archive for Space Telescopes at the Space Telescope Science Institute, which is operated by the Association of Universities for Research in Astronomy, Inc., under NASA contract NAS 5-03127 for JWST. These observations are associated with program \#03181. Support for program \#03181 was provided by NASA through a grant from the Space Telescope Science Institute, which is operated by the Association of Universities for Research in Astronomy, Inc., under NASA contract NAS 5-03127. YZ acknowledges support from a Heising-Simons Foundation 51 Pegasi b Alumni Faculty Grant (\#2023-4808).

We thank the referee for their careful attention and critiques which were very helpful in refining our analysis and presentation.
\end{acknowledgements}

\facilities{JWST (NIRSpec)}

\software{astropy \citep{ast13,astropy2018,astropy2022}, matplotlib \citep{hun07}, numpy \citep{van11,harris2020array}, octofitter \citep{Hoffman2014,Xu2020,Thompson2023}, petitRADTRANS \citep{pRT,Nasedkin2024a,Lei2024}, scipy \citep{jon01,scipy_curvefit_source,2020SciPy-NMeth}, xarray \citep{xarray}}

\end{document}